\documentclass[12pt]{article}

\usepackage{algpseudocode}
\usepackage{amsmath, amssymb, amsthm}
\usepackage{float,graphicx, multirow,parskip, subcaption, setspace}
\usepackage{comment}
\usepackage{url}
\usepackage{enumitem}
\usepackage{rotating}
\usepackage{xr, xr-hyper}
\usepackage{hyperref}

\usepackage{dsfont}
\usepackage[scr=boondox]{mathalfa}

\usepackage{natbib}
\usepackage{placeins}
\usepackage{algorithm2e}
\RestyleAlgo{ruled}

\newcommand{\bx}{\boldsymbol{x}}

\newcommand{\N}{\mathcal{N}}
\def\P{\mathbb{P}}
\newcommand{\E}{\mathbb{E}}

\newcommand{\tto}{\textrm{TTO}}
\newcommand{\ttos}{\textrm{TTOs}}
\newcommand{\ttoone}{\textrm{1TTO}}
\newcommand{\ttotwo}{\textrm{2TTO}}
\newcommand{\ttothree}{\textrm{3TTO}}
\newcommand{\ttop}{\textrm{TTOP}}
\newcommand{\ttops}{\textrm{TTOPs}}
\newcommand{\ttoptwo}{\textrm{2TTOP}}
\newcommand{\ttopthree}{\textrm{3TTOP}}
\newcommand{\xwoba}{\textrm{xwOBA}}
\newcommand{\woba}{\textrm{wOBA}}

\newcommand{\ind}[1]{\mathbb{I}\left(#1\right)}

\usepackage{fullpage, parskip}
\title{A Bayesian analysis of the time through the order penalty in baseball}
\author{Ryan S. Brill\thanks{Graduate Group in Applied Mathematics and Computational Science, University of Pennsylvania. Correspondence to: ryguy123@sas.upenn.edu}, Sameer K. Deshpande\thanks{Dept.~of Statistics, University of Wisconsin--Madison}, and Abraham J. Wyner\thanks{Dept.~of Statistics and Data Science, The Wharton School, University of Pennsylvania}}

\begin{document}
\maketitle
% abstract
\begin{abstract}
As a baseball game progresses, batters appear to perform better the more times they face a particular pitcher.
% The apparent drop-off in pitcher performance averaged across each time through the order, known as a Time Through the Order Penalty ($\ttop$), is often attributed to within-game batter learning.
The apparent drop-off in pitcher performance from one time through the order to the next, known as the Time Through the Order Penalty ($\ttop$), is often attributed to within-game batter learning.
Although the $\ttop$ has largely been accepted within baseball and influences many managers' in-game decision making, we argue that existing approaches of estimating the size of the $\ttop$ cannot disentangle continuous evolution in pitcher performance over the course of the game from discontinuities between successive times through the order.
% the discontinuous change in pitcher performance from one time through the order to the next from the continuous change in pitcher decline across all the opposing batters.
% batter learning from pitcher fatigue. 
% Using a Bayesian multinomial regression model, we show that, after adjusting for confounders like batter and pitcher quality, handedness, and home field advantage, pitcher fatigue is the main driver of the $\ttop$.
% the $\ttop$ is actually an artifact of continuous pitcher decline.
Using a Bayesian multinomial regression model, we find that, after adjusting for confounders like batter and pitcher quality, handedness, and home field advantage, there is little evidence of strong discontinuity in pitcher performance between times through the order.
% there is little evidence of a strong batter learning effect.
% We specifically show that expected weighted on-base average increases steadily over the course of the game and does not display sharp discontinuities reflecting substantial batter learning between times through the order.  
Our analysis suggests that the start of the third time through the order should not be viewed as a special cutoff point in deciding whether to pull a starting pitcher. 
\end{abstract}

\newpage
\section{Introduction}
\label{sec:introduction}

%%%%%%%%%%%%%%%%%%%%%%%%%%%%%%%%%%%%%%%%%%%%%%%%%%%%%%%%%%%%%%%%%%%%%%%%%%%%%%%%%%%%%%%%%%%%%%%%%%%%%%%%%%

In Game 6 of the 2020 World Series, the Tampa Bay Rays' manager, Kevin Cash, pulled his starting pitcher, Blake Snell, midway through the sixth inning.
When he was pulled, Snell had been pitching extremely well; he had allowed just two hits and struck out nine batters on 73 pitches.
Moreover, the Rays had a one run lead.
Snell's replacement, Nick Anderson, promptly gave up two runs, which ultimately proved decisive: the Rays went on to lose the game and the World Series.
After the game, Cash justified his decision to pull Snell, remarking that he ``didn't want Mookie [Betts] or [Corey] Seager seeing Blake a third time'' \citep{rivera20}. 

In his justification, Cash cites the third \textit{Time Through the Order Penalty} ($\ttop$), which was first formally identified in \citet[pp.~187--190]{theBook} and recently popularized by \citet{bballProspectusTTO}.
It has long been observed that, on average, batters tend to perform better the more times they face a pitcher; for instance, they tend to get on base more often on their third time facing a pitcher than their second.
\citet{theBook} quantified the corresponding drop-off in pitcher performance as increases in \textit{weighted on-base average} ($\woba$; see Section~\ref{sec:wOBA} for details).
They observed that the average $\woba$ of a plate appearance in the first time through the order ($\ttoone$) is about 9 $\woba$ points less than that in the second $\tto$ ($\ttotwo$).
Further, the average $\woba$ of a plate appearance in the second $\tto$ is about 8 $\woba$ points less than that in the third $\tto$ ($\ttothree$) \citep[Table 81]{theBook}.

The $\ttop$ is considered canon by much of the baseball community.
Announcers routinely mention the $\ttopthree$ during broadcasts and several managers regularly use the $\ttopthree$ to justify their decisions to pull starting pitchers at the start of the third $\tto$.
For instance, A.J. Hinch, who managed the Houston Astros from 2015 to 2019, noted ``the third time through is very difficult for a certain caliber of pitchers to get through.''
Brad Ausmus, who managed the Detroit Tigers from 2014 to 2017, explained ``the more times a hitter sees a pitcher, the more success that hitter is going to have'' \citep{MLBmanagersTTOPquotes}.

\citet{theBook} attribute the increased average $\woba$ from one $\tto$ to the next to within-game batter learning.
According to them, batters learn the tendencies of a pitcher as the game progresses.
In fact, they observe ``pitchers hitting a wall after 10 or 11 batters'' rather than a ``steady degradation in [pitcher] performance'' \citep[pg. 189]{theBook}. 
\citet{bballProspectusTTO} agrees and goes further, stating ``the $\ttop$ is not about fatigue. It is about [batter] familiarity.''

We argue that \citet{theBook}'s analysis is insufficient to justify such sweeping conclusions.
\citet{theBook} estimated the $\ttoptwo$ and $\ttopthree$ by first binning plate appearances by lineup position and $\tto$.
They then computed the average $\woba$ within each bin. 
% Their analysis, by design, cannot disentangle batter learning from pitcher fatigue. 
% More generally, their analysis cannot disentangle continuous evolution in pitcher peformance over the course of a game from discontinuities in pitcher performance from each $\tto$ to the next.
Their analysis, by design, cannot disentangle continuous evolution in pitcher performance over the course of a game (e.g., from pitcher fatigue) from discontinuities between successive $\ttos$ (e.g., from batter learning).
Further, they provide no uncertainty quantification about their estimated $\ttops.$ 

We conduct a more rigorous statistical analysis of the trajectory of pitcher performance over the course of a baseball game. 
Specifically, we fit a Bayesian multinomial logistic regression model to predict the outcome of each plate appearance as a function of the \textit{batter sequence number}, batter quality, pitcher quality, handedness match, and home field advantage. 
The batter sequence number simply counts how many batters the pitcher has faced up to and including the current plate appearance. 
% We find that the expected $\woba$ forecasted by our model increases steadily over the course of a game and does not display the sharp discontinuities that would arise from systematic batter learning between times through the order.
We find that the expected $\woba$ forecast by our model increases steadily over the course of a game and does not display sharp discontinuities between times through the order.
% In other words, we find little evidence to attribute the $\ttop$ to batter learning.
Based on these results, we recommend managers cease pulling starting pitchers at the beginning of the $\ttothree$.

The remainder of this paper is organized as follows.
We introduce our Bayesian multinomial logistic regression model of a plate appearance outcome in Section~\ref{sec:model_specification}. 
% In Section~\ref{sec:model_sim}, we run a simulation study that demonstrates our proposed model has the capacity to detect a $\ttop$. 
We present our main findings in Section~\ref{sec:results} and conclude by discussing implications of our results in Section~\ref{sec:baseball_implications}.

% %%%%%% REFERENCES
% \clearpage
% \bibliography{../refs}
% \bibliographystyle{plainnat}

\section{Data and model specification}
\label{sec:model_specification}
% %%%%%%%%%%%%%%%%%%%%%%%%%%%%%
% \usepackage{fullpage, parskip}
% \onehalfspacing
% %%%%%%%%%%%%%%%%%%%%%%%%%%%%%

% \label{sec:model_specification}

We begin with a brief overview of our MLB plate appearance dataset and identify several variables that may be predictive of the outcome of a plate appearance. 
We then introduce our Bayesian multinomial logistic regression model.

%%%%%%%%%%%%%%%%%%%%%%%%%%%%%%%%%%%%%%%%%%%%%%%%%%%%%%%%%%%%%%%%%%%%%%%%%%%%%%%%%%%%%%%
\subsection{Retrosheet data}\label{sec:theData}
 
%We relied on play-by-play data from Retrosheet in our analysis.
We scraped every plate appearance from 1990 to 2020 from the Retrosheet database.
For each plate appearance, we record the outcome (e.g.~out, single, etc.), the event $\woba$, the handedness match between the batter and pitcher, and whether the batter is at home.
We further compute measures of batter and pitcher quality for each plate appearance (see Section~\ref{sec:quality} for details).
We include our final dataset, along with all pre-processing and data analysis scripts in the Supplementary Materials.
We used \textsf{R} \citep{R} for all analyses. 

We restrict our analysis to every plate appearance from 2012 to 2019 featuring a starting pitcher in one of the first three times through the order, using the 2017 season as our primary example. 
We remove plate appearances featuring switch hitters from our dataset.
Our 2017 dataset consists of $108,519$ plate appearances, 691 unique batters, and 315 unique starting pitchers. 

There are $K = 7$ possible outcomes of a plate appearance: out, unintentional walk (uBB), hit by pitch (HBP), single (1B), double (2B), triple (3B), and home run (HR). 
For each $i = 1, \ldots, n$, let $y_{i}$ be the categorical variable indicating the outcome of the $i^{\text{th}}$ plate appearance. 
Notationally, we write
\begin{equation}
y_i \in \{1,2,...,7\} = \{\text{Out, uBB, HBP, 1B, 2B, 3B, HR}\}.
\label{eqn:wOBAcategories}
\end{equation}

In predicting the probability of each plate appearance outcome, we need to control for several factors.
We introduce the \textit{batter sequence number} $t \in \{1,...,27\}$, which records how many batters the pitcher has faced up to and including that plate appearance.
We additionally construct indicators of being in the $\ttotwo$ and $\ttothree$, $\ind{10 \leq t \leq 18}$ and $\ind{19 \leq t \leq 27}.$
% Finally, we consider $\texttt{hand},$ an indicator that is equal to one when the batter and pitcher have matching handedness and zero otherwise, and $\texttt{home},$ an indicator that is equal to one when the batter is at home and zero otherwise.

Intuitively, we expect that most pitchers are more likely to give up base hits and home runs to elite batters than they are to strike out elite batters. 
Similarly, we expect elite pitchers would have more plate appearances ending in outs than base hits against most batters.
Accordingly, when modeling the outcome of a plate appearance, we adjust for the quality or skill of the batter and pitcher.
% To this end, let $x^{(p)}$ and $x^{(b)}$ be the logit-transformed estimates of pitcher and batter quality, respectively.
To this end, let $x^{(p)}$ and $x^{(b)}$ denote the estimates of pitcher and batter quality, respectively.
We discuss the computation of both quality measures in Section~\ref{sec:quality}.

Additionally, we expect that a pitcher whose handedness matches that of the batter (e.g., the pitcher and batter are both right handed) is less likely to give up base hits and home runs than a pitcher whose handedness doesn't match the batter's.
To this end, we define $\texttt{hand}$, an indicator that is equal to one when the batter and pitcher have matching handedness and zero otherwise.
Finally, we expect that a pitcher on the road is more likely to give up base hits and home runs than a pitcher at home. 
Thus we define $\texttt{home}$, an indicator that is equal to one when the batter is at home and zero otherwise.

Table \ref{table:covariateDescriptions} summarizes the variables that we record from plate appearance $i.$ 
\begin{table}[h]
\centering
\caption{Summary of variables measured for each at-bat that are used in our analysis.}
\label{table:covariateDescriptions}
\begin{tabular}{ll} \hline
Covariate symbol & Covariate description \\ \hline
$y_i$ & {outcome of the $i^{\text{th}}$ plate appearance $\in \{1,...,K=7\}$} \\ 
$t_i$  & {the batter sequence number $\in \{1,...,27\}$} \\ 
$\ind{t_i \in \ttotwo}$ & {binary variable indicating whether the pitcher is in his second $\tto$} \\ 
$\ind{t_i \in \ttothree}$  & {binary variable indicating whether the pitcher is in his third $\tto$} \\ 
$x^{(b)}_i$ & {running-average estimator of batter quality} \\ 
$x^{(p)}_i$ & {running-average estimator of pitcher quality} \\ 
% $x^{(b)}_i$ & {logit-transformed running-average estimator of batter quality} \\ 
% $x^{(p)}_i$ & {logit-transformed running-average estimator of pitcher quality} \\ 
$\texttt{hand}_i$ & {binary variable indicating handedness match between batter and pitcher} \\ 
$\texttt{home}_i$ & {binary variable indicating whether the batter is at home} \\ 
$\bx_{i}$  & {$\bx_{i}  = ( x^{(b)}_i, \ x^{(p)}_i, \ \texttt{hand}_i, \ \texttt{home}_i )$} \\
%$\bx_{i}$  & {$\bx_{i}  = [\text{logit}( x^{(b)}_i), \ \text{logit}( x^{(p)}_i), \ \texttt{hand}_i, \ \texttt{home}_i]$} \\
\hline
\end{tabular}
\end{table}

%%%%%%%%%%%%%%%%%%%%%%%%%%%%%%%%%%%%%%%%%%%%%%%%%%%%%%%%%%%%%%%%%%%%%%%%%%%%%%%%%%%%%%
\subsection{A multinomial logistic regression model}
\label{sec:model}

% Pitcher fatigue and batter learning are two potential mechanisms for the decline in pitcher performance over the course of the game.
% We explicitly account for both mechanisms in our model of each plate appearance outcome.
We fit a Bayesian multinomial logistic regression model to predict the outcome of each plate appearance.
For each non-out result $(k \neq 1)$, we model
\begin{align}
\label{eqn:model}
\log\left(\frac{\P(y_{i} = k)}{\P(y_{i} = 1)}\right) &= \alpha_{0k} + \alpha_{1k}t_{i} + \beta_{2k}\ind{t_{i} \in \ttotwo} + \beta_{3k}\ind{t_{i} \in \ttothree} + \bx_{i}^{\top}\eta_{k},
\end{align}
where the vector $\bx_{i}$ concatenates our batter and pitcher quality and indicators for handedness and home team: $\bx_{i}^{\top} = (x_{i}^{(b)}, x_{i}^{(p)}, \texttt{hand}_{i}, \texttt{home}_{i}).$
% Recall that $x_{i}^{(b)}$ and $x_{i}^{(p)}$ are the logit-transformed estimates of batter and pitcher quality.
% We felt it was more natural to allow the log-odds of each plate appearance outcome to evolve non-linearly with respect to these quality metrics.
% Specifically, we find it plausible that there are diminishing returns at both extremes of player quality.
% That is, we did not expect a small change in pitcher quality to manifest the same changes in the log-odds of a particular plate appearance outcome for a mediocre pitcher, an average pitcher, or an elite pitcher (keeping all else constant).
% The logit transformation allows us to capture this phenomenon.
% While this choice may appear somewhat unusual, we have found that it also yields a model with better predictive accuracy than a model that uses the raw quality covariates (see Appendix~\ref{sec:model_performance}).

% In Model~\eqref{eqn:model}, the parameters
The parameters $\alpha_{0k}$ and $\alpha_{1k}$ control the continuous evolution of the probability of each plate appearance outcome throughout the game.
In contrast, the parameters $\beta_{2k}$ and $\beta_{3k}$ allow for discontinuities in these probabilities between different times through the order.
% Accordingly, we interpret the term $\alpha_{0k} + \alpha_{1k}t$ as the continuous effect of pitcher decline on the probabilities of each outcome.
Pitchers face each of the opposing team's batters, and so we interpret the term $\alpha_{0k} + \alpha_{1k}t$ as the \textit{continuous} effect of a change in pitcher performance on the probability of each outcome.
Batters, on the other hand, take turns facing the opposing team's pitcher, and so we interpret $\beta_{2k}$ and $\beta_{3k} - \beta_{2k}$ as the respective \textit{discontinuous} effects of a change in batter performance between the first and second times through the order and between the second and third times through the order.
Observe that for $k \neq 1,$ a large positive value of $\beta_{2k}$ suggest that the non-out outcome $k$ is systematically more likely to occur in the the second time through the order than the first.
Similarly, a large positive positive value of $\beta_{3k} - \beta_{2k}$ suggests that the outcome is more likely to occur in the third time through the order than the second.
Consequently, based on our model parametrization, we would anticipate the $\ttoptwo$ and $\ttopthree$ to manifest as positive values $\beta_{2k}$ and $\beta_{3k} - \beta_{2k}.$

Our model allows the log-odds of each non-out plate appearance outcome to evolve linearly with batter sequence number.
A more flexible model wouldn't enforce a particular functional form on the change in pitcher performance over the course of a game.
Additionally, our model assumes that the trajectory of within-game pitcher deterioration is the same across all pitchers and batters.
A more elaborate model would allow within-game performance to change at at different rates for different players.
We find that using these more elaborate models doesn't change the qualitative results of our study (see Appendix~\ref{app:alt_models}). 

% Previous research suggests 
Moreover, previous research suggests
that pitchers decline continuously over the course of the game; \citet{greenhouse13}, for instance, documented continuous decreases in pitch velocity.
On this view, the longer a pitcher stays in the game, the more likely he is to allow non-out outcomes in a plate appearance due to his continuous deterioration.
We encode our intuition in Model~\eqref{eqn:model} by constraining the slopes $\alpha_{1k}$ to be positive with a truncated prior:
\begin{equation}
\label{eqn:slope_prior}
\alpha_{1k} \sim \text{half-}t_7.
\end{equation}
We specify standard normal priors to all of our other coefficients,
\begin{equation}
\label{eqn:priors}
\alpha_{0k}, \ \beta_{2k}, \ \beta_{3k}, \ \eta_{\ell k} \sim \mathcal{N}(0,1).
\end{equation}
Note that the qualitative results of our study remain the same when we use a diffuse prior $\mathcal{N}(0,25)$ and drop the positive-slope constraint (see Appendix~\ref{sec:varying_slopes_model}).

Because the posterior distribution of $(\alpha, \beta, \eta)$ is not analytically tractable, we use Markov Chain Monte Carlo (MCMC) to draw approximate samples from the posterior distribution.
We implement our sampler in \textsf{Stan} \citep{Stan} and perform our MCMC simulation using the \textbf{rstan} package \citep{rstan}.
We use a high-performance computing cluster to run all of our computations.
%All of our computations are run on a high-performance computing cluster \citep{whartonhpcc}.

Additionally, in Appendix~\ref{sec:model_sim} we conduct a simulation study to assess the capacity of our model to estimate time through the order penalties of various sizes.
Specifically, we simulate data consistent with different $\ttops$ and verify that our posterior estimates are close to the data generating parameters.

%%%%%%%%%%%%%%%%%%%%%%%%%%%%%%%%%%%%%%%%%%%%%%%%%%%%%%%%%%%%%%%%%%%%%%%%%%%%%%%%%%%%%%%%%
\subsection{Selection bias}\label{sec:selection_bias_adjustment} 

\begin{figure}[h]
\centering
\includegraphics[width = \textwidth]{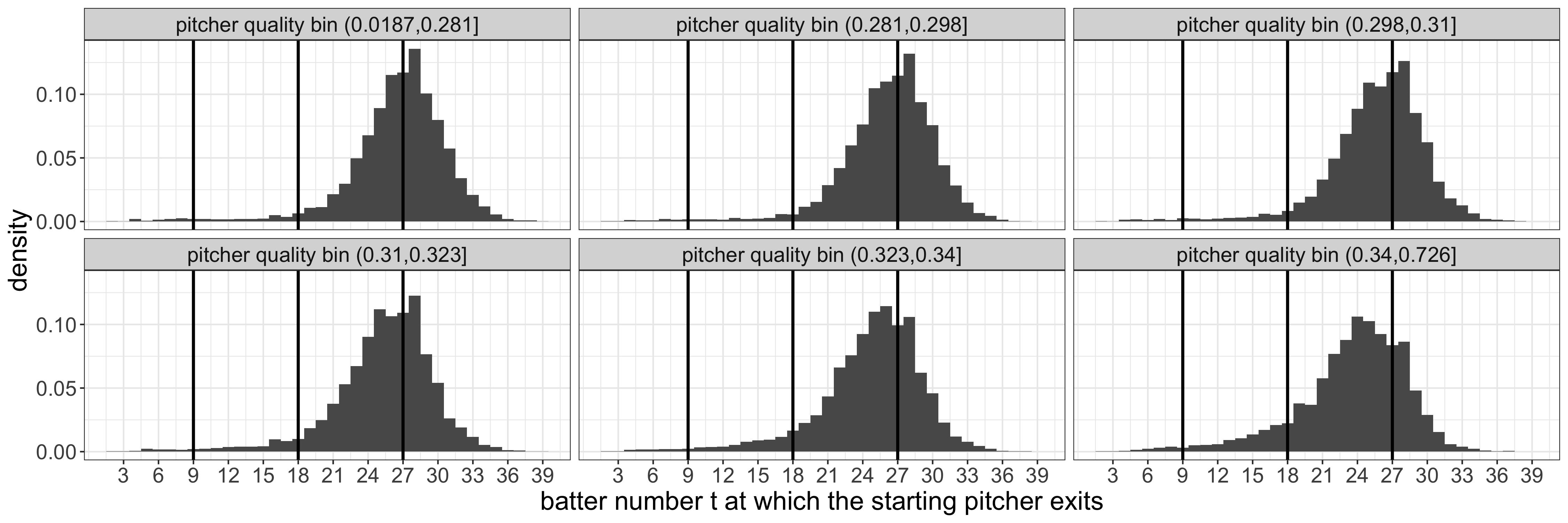}
\caption{
Histogram of the batter sequence number $t$ at which a starting pitcher exits the game 
for different bins of pitcher quality.
% for six quantile bins of pitcher quality.
}
\label{fig:visualize_selection_bias1}
\end{figure}

We are primarily interested in understanding the magnitude and significance of discontinuous pitcher decline between times through the order.
Formally, we are interested in the parameters $\beta_{2k}$ and $\beta_{3k}$ from our Model~\eqref{eqn:model}.
Ideally we want to estimate $\beta$ in the counterfactual scenario in which each pitcher faces each of the first 27 opposing batters.
But, we cannot conduct a randomized controlled experiment; rather, we use observational data which is subject to the selection process of a baseball manager removing his starting pitcher.
% In Figure~\ref{fig:visualize_selection_bias1} we visualize this selection process.
% % In particular, we see that low quality pitchers, from the pitcher quality bins $(0.323,0.34]$ and $(0.34,0.726]$, are sometimes removed in $\ttotwo$, whereas other pitchers are almost always removed in $\ttothree$.
% % In other words, the pitchers who remain in the game in $\ttothree$ have better pitcher quality over average than those who remain in the game in $\ttotwo$.
% % Note that the six pitcher quality bins in Figure~\ref{fig:visualize_selection_bias1} are six evenly sized quantiles of pitcher quality.
% In particular, we see that low quality pitchers, from the pitcher quality bin $(0.34,0.726]$, are removed slightly earlier in the game than better pitchers, as the corresponding histogram is shifted slightly to the left.
% % In other words, the pitchers who remain in the game in $\ttothree$ have better pitcher quality over average than those who remain in the game in $\ttotwo$.
% Note that the six pitcher quality bins in Figure~\ref{fig:visualize_selection_bias1} are six evenly sized quantiles of pitcher quality.
We visualize this selection process in Figure~\ref{fig:visualize_selection_bias1}, which shows that worse pitchers (pitcher quality larger than, say, $0.34$) are slightly more likely to be removed earlier in the game, as the corresponding histogram is shifted slightly to the left.
Note that the six pitcher quality bins in Figure~\ref{fig:visualize_selection_bias1} are six evenly sized quantiles of pitcher quality.

Because our dataset is missing some $\ttothree$ batting observations against worse pitchers, fitting Model~\eqref{eqn:model} on our raw dataset may lead to a lower estimated probability of each non-out plate appearance outcome in $\ttothree$.
To combat this, we remove all games from our dataset in which the starting pitcher is pulled prior to $\ttothree$.
In 2017, for instance, this reduces our dataset by $8\%$ from $4860$ games to $4469$ games.
Then, we fit our Model~\eqref{eqn:model} on the reduced dataset, and we interpret our results as a $\ttop$ (or lack thereof) \textit{conditional on getting through $\ttotwo$}.
Conditional on getting through $\ttotwo$, our dataset of all starting pitcher at-bats in the first three $\ttos$ is balanced on the pitcher quality covariate, and so the $\ttop$ discontinuity parameters $\beta$ won't be biased due to the selection process.
In other words, after our data truncation, the distribution of pitcher quality is similar for each batter sequence number $t$ .

Even after truncating our dataset, since most starting pitchers are removed during $\ttothree$, our dataset is missing observations at the end of $\ttothree$.
If each pitcher were allowed to pitch to the end of $\ttothree$, it is plausible that he would perform even worse than he did earlier in the game due to, for instance, additional fatigue.
Therefore, we still underestimate the continuous pitcher decline parameters $\alpha$.
Nonetheless, as we are primarily interested in the discontinuity parameters $\beta$ and not the continuity parameters $\alpha$, we leave a more elaborate estimation of continuous pitcher decline to future work.

%%%%%%%%%%%%%%%%%%%%%
%%%%%%%%%%%%%%%%%%%%%

% %%%%%%%%%%%%%%%%%%%%%
% \begin{figure}[h]
% \begin{subfigure}{\textwidth}
% \centering
% \includegraphics[width = \textwidth]{plots/selection_bias_explore_plot_2.png}
% \caption{}
% \label{fig:visualize_selection_bias1}
% \end{subfigure}
% \\
% \begin{subfigure}{\textwidth}
% \centering
% \includegraphics[width = \textwidth]{plots/selection_bias_explore_plot_5.png}
% \caption{}
% \label{fig:visualize_selection_bias2}
% \end{subfigure}
%     \caption{Histogram of the batter sequence number $t$ at which a starting pitcher exits the game, for different bins of pitcher quality (a) and different bins of mean game $\woba$ (b).}%
%     \label{fig:visualize_selection_bias}
% \end{figure}
% %%%%%%%%%%%%%%%%%%%%%

%%%%%%%%%%%%%%%%%%%%%%%%%%%%%%%%%%%%%%%%%%%%%%%%%%%%%%%%%%%%%%%%%%%%%%%%%%%%%%%%%%%%%%%%%
\subsection{Measuring pitcher performance via $\woba$}\label{sec:wOBA} 

\textbf{Weighted on-base average.}
Although Model~\eqref{eqn:model} allows us to examine potential $\ttops$ for each plate appearance outcome, such multivariate measures are somewhat difficult to interpret and compare.
We instead focus on quantifying the $\ttop$ using a much more interpretable quantity, \textit{weighted on-base average} ($\woba$), which was first introduced in \citet{theBook}.
%We consider 7 possible outcomes for a plate appearance: out, unintentional walk (uBB), hit by pitch (HBP), single (1B), double (2B), triple (3B), and home run (HR). 
%Although it is possible to quantify pitcher and batter performance in terms of the probabilities of each of these outcomes, such multivariate measures are somewhat difficult to interpret and compare.
%Hence we also quantify pitcher and batter performance on a more naturally interpretable scale: \textit{weighted on-base average} ($\woba$), which was originally formulated in \citet{theBook}. 

$\woba$ overcomes many limitations of traditional metrics like batting average, on-base percentage, and slugging percentage. 
Briefly, batting average and on-base percentage treat all hits equally, with singles being worth as much as triples.
Slugging percentage attempts to reward different types of hits differently, but does so in too simplistic of a fashion: in computing slugging percentage, a triple is worth three times what a single is worth. 
Such weighting is arbitrary, and is not tied to the relative impact of a triple over a single with regard to, say, run scoring or win probability.
$\woba$ combines the different aspects of offensive production into one metric, weighing each offensive action in proportion to its actual run value \citep{wOBAExplanation}. 
The $\woba$ of a plate appearance is simply the weight associated with the offensive action of the outcome. 
Specifically, the 2019 $\woba$ weight of each offensive action in decreasing order is 1.940 for a home run (HR), 1.529 for a triple (3B), 1.217 for a double (2B), 0.870 for a single (1B), 0.719 for hit-by-pitch (HBP), 0.690 for unintentional walks (uBB), and 0 for an out (OUT) \citep{wobaCharts}. 
$\woba$ is rescaled so that the league average $\woba$ equals the league average on-base percentage. Throughout this paper, we use 2019 $\woba$ weights for each season. Additionally, we usually refer to \textit{$\woba$ points}, which is $\woba$ multiplied by 1000, to be consistent with the baseball community's use of $\woba$. 

\begin{figure}[h]

\begin{subfigure}{0.49\textwidth}
\centering
\includegraphics[width = \textwidth]{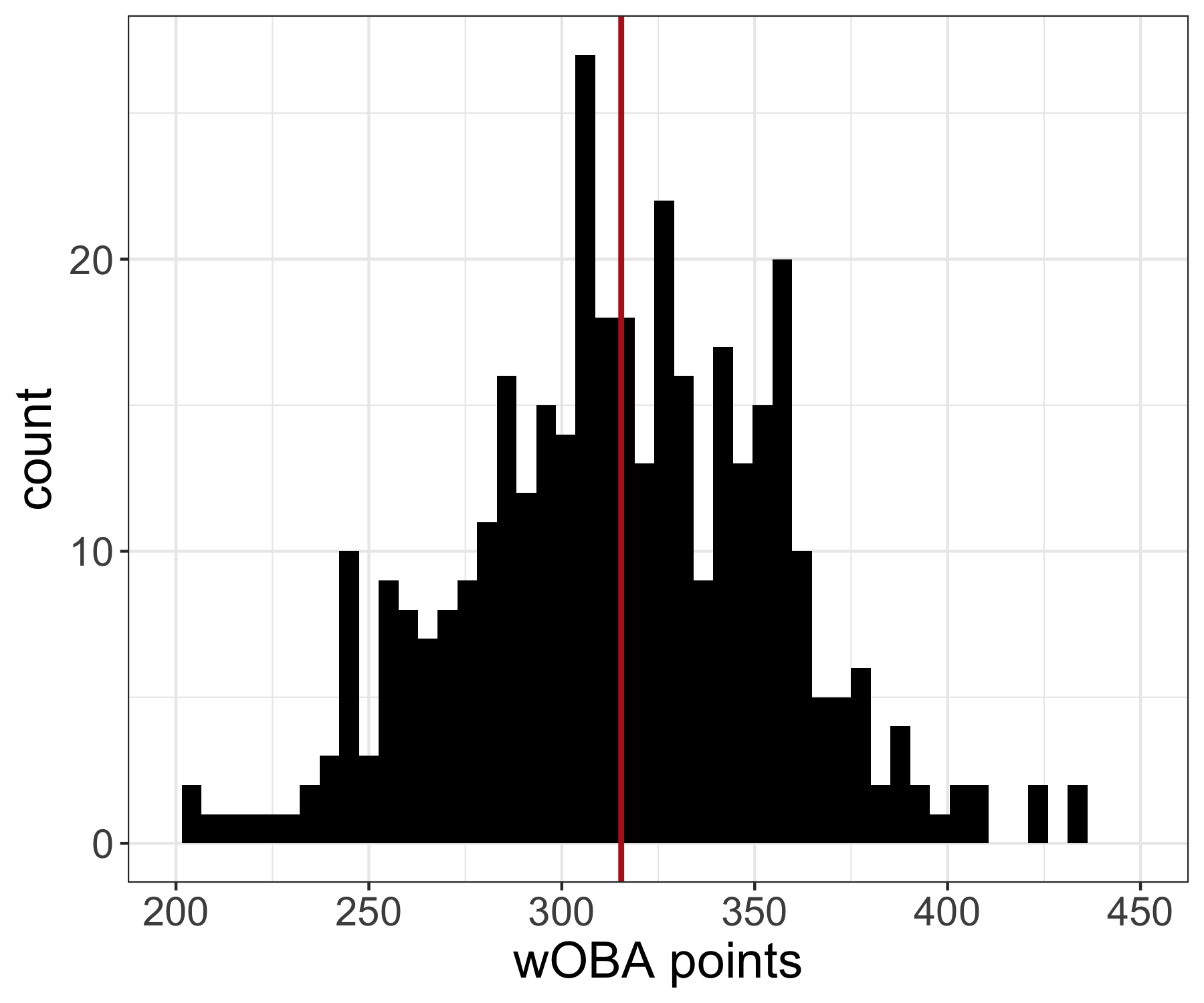}
\caption{}
\label{fig:bat_woba}
\end{subfigure}
\begin{subfigure}{0.49\textwidth}
\centering
\includegraphics[width = \textwidth]{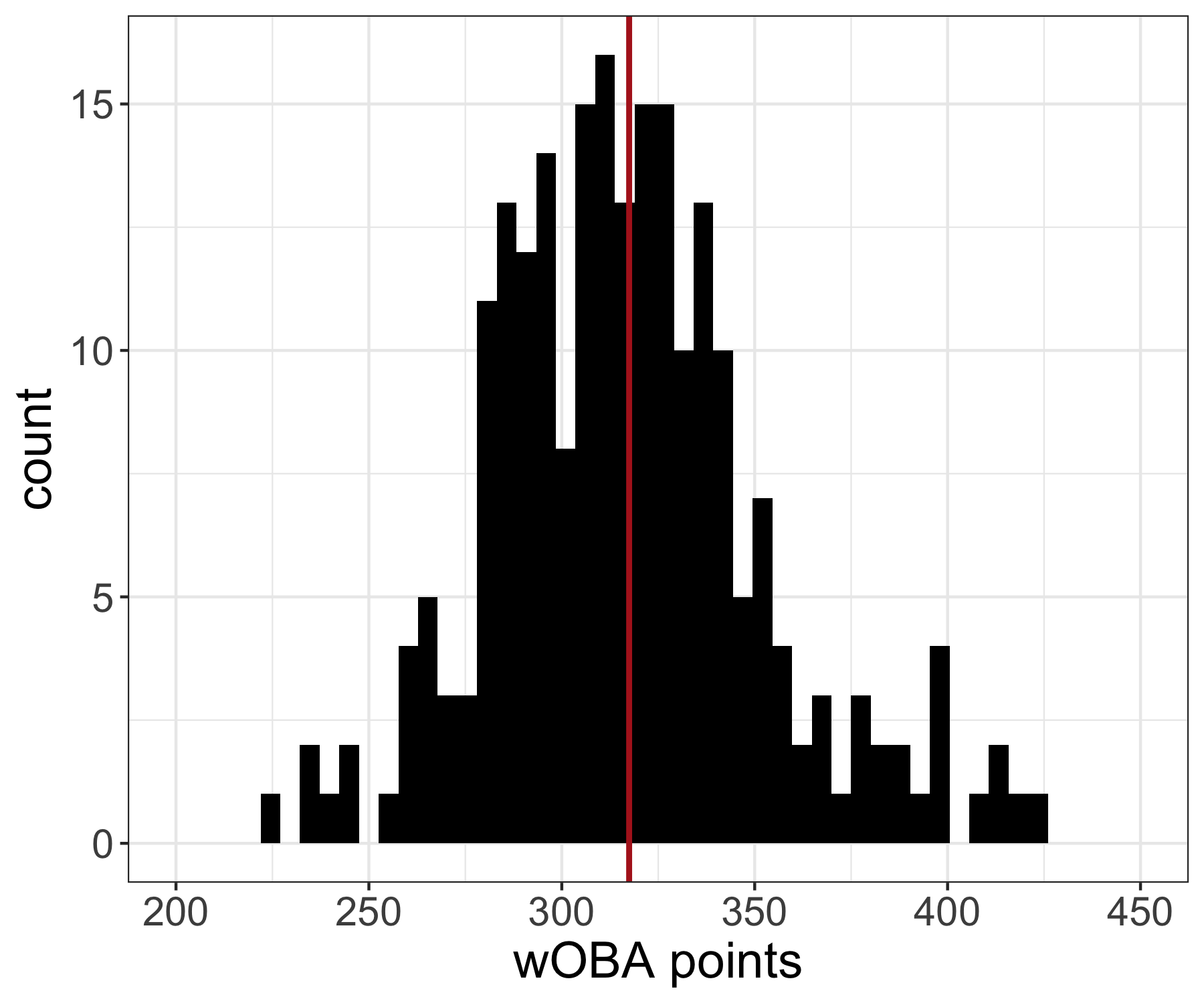}
\caption{}
\label{fig:pit_woba}
\end{subfigure}
    \caption{The distribution of end-of-season mean $\woba$ points for all batters in 2017 (a) and all pitchers in 2017 (b) with over 100 plate appearances, using 2019 $\woba$ weights. The red line denotes the mean.}%
    \label{fig:woba_dists}%
\end{figure}

To understand the effect size of a potential time through the order penalty, it is important to understand the distribution of $\woba$ points across batters and pitchers. 
In Figure \ref{fig:woba_dists}, we plot the distribution of end-of-season mean plate appearance $\woba$ points for all batters and for all pitchers in 2017 who have over 100 plate appearances. Both batters and pitchers have a median $\woba$ points of 315. 
The standard deviation of $\woba$ points for batters is 41.5, and for pitchers is 36.7. 

%%%%%%%%%%%%%%%%%
%%%%%%%%%%%%%%%%%

\textbf{Expected weighted on-base average.}
Using Model~\eqref{eqn:model}, we can predict the probability of each plate appearance outcome.
We can use these predicted probabilities to derive an \textit{expected $\woba$} for each plate appearance.
We use expected $\woba$ to examine the trajectory of pitcher performance throughout the game.
%Furthermore, we summarize the posterior distribution of Model~\ref{eqn:model} by examining the trajectory of pitcher performance, via \textit{expected $\woba$}, over the course of the game.

To this end, let $k \in \{1,...,K\}$ denote the outcome of a plate appearance and let $t \in \{1,...,27\}$ denote the $t^{th}$ batter a pitcher faces in a game. 
Also, let $x^{(b)}$ be the logit-transformed quality of the batter, $x^{(p)}$ the logit-transformed quality of the pitcher, \texttt{hand} be the binary indicator of the handedness match between the batter and pitcher, and \texttt{home} be the binary indicator of home field advantage. 
Define the \textit{plate-appearance-state vector} $\bx$ by
\begin{equation}
\label{eqn:x_ex1}
\bx^\top = ( x^{(b)},x^{(p)}, \texttt{hand}, \texttt{home}).
\end{equation}
Then, according to Model~\eqref{eqn:model}, the probability that a plate appearance involving the $t^{th}$ batter of a game and plate-appearance-state vector $\bx$ results in outcome $k$ is
\begin{equation}
\label{eqn:prob_ktx}
 \P(y = k|t,\bx) = \frac{\lambda_k(t,\bx)}{\sum_{j=1}^{K} \lambda_j(t,\bx) },
\end{equation}
where 
\begin{equation}
\lambda_k(t,\bx) = \exp\big(\alpha_{0k} + \alpha_{1k} t
 + \beta_{2k}\ind{t \in \ttotwo}
 + \beta_{3k}\ind{t \in \ttothree}
 + \bx^\top \eta_{k}\big) 
\end{equation}
when $k \neq 1$ and $\lambda_k(t,\bx) = 1$ when $k=1$.
From this, we define the \textit{expected $\woba$ points} of a plate appearance involving the $t^{th}$ batter of a game and plate-appearance-state vector $\bx$ by
\begin{equation}
\label{eqn:xwOBA_tx}
\xwoba(t,\bx) = \sum_{k=1}^{K} 1000 \cdot w_k \cdot \P(y = k|t,\bx),
\end{equation}
where $w_k$ is the $\woba$ weight of the $k^{th}$ plate appearance outcome. 

% To visualize the nature of within-game pitcher decline implied by Model~\eqref{eqn:model}, we in Sections~\ref{sec:model_sim} and \ref{sec:results} plot the trajectory of the expected $\woba$ of a plate appearance over the course of a game, 
To visualize the nature of within-game pitcher decline implied by Model~\eqref{eqn:model}, we in Section~\ref{sec:results} plot the trajectory of the expected $\woba$ of a plate appearance over the course of a game, 
\begin{equation}
\label{eqn:xwoba_seq}
\big\{ \xwoba(t,\bx) \big\}_{t=1}^{27},
\end{equation}
holding the plate-appearance-state vector $\bx$ constant.

%%%%%%%%%%%%%%%%%%%%%%%%%%%%%%%%%%%%%%%%%%%%%%%%%%%%%%%%%%%%%%%%%%%%%%%%%%%%%%%%%%%%%%%%%

\subsection{Definitions of pitcher and batter quality}
\label{sec:quality}

% % To measure batter quality, we could use a batter's end-of-season average $\woba$. 
% % Doing so, however, introduces a form of \textit{data bleed} into our analysis: the $\woba$ of the $i^{\text{th}}$ plate appearance $y_i$ is used to compute the batter's end-of-season average $\woba$ so to use it as a covariate is to use $y_i$ to help predict $y_{i}.$ 
% % Moreover, to use a batter's end-of-season average $\woba$ as a covariate is to use future information to predict the outcome of a plate appearance. 
% To measure batter quality, we could use a batter's end-of-season average $\woba$. 
% Doing so, however, introduces a form of \textit{data bleed} into our analysis: 
% % $y_i$, the $\woba$ of the $i^{\text{th}}$ plate appearance, is used to compute the batter's end-of-season average $\woba$,
% $y_i$, the $\woba$ of the $i^{\text{th}}$ plate appearance, is used to compute the batter's end-of-season average $\woba$, so to use it as a covariate is to use $y_i$ to help predict $y_{i}.$ 
% % Moreover, to use a batter's end-of-season average $\woba$ as a covariate is to use future information to predict the outcome of a plate appearance. 

To measure batter quality, we could use a batter's end-of-season average $\woba$. 
Doing so, however, introduces a form of \textit{data bleed} into our analysis: $y_i$, the $\woba$ of the $i^{\text{th}}$ plate appearance, is used to compute the batter's end-of-season average $\woba$, so to use it as a covariate is to use $y_i$ to help predict $y_{i}.$ 
To avoid data bleed, we could instead use a batter's average $\woba$ over all prior plate appearances during the current season. 
Early in the season, however, this metric is extremely noisy. 
Hence we introduce a normal-normal conjugate running-average estimator that early in the season is close to a batter's average $\woba$ from the end of his previous season and that is closer to his current average $\woba$ later in the season. 

Specifically, let $ x_{bj}$ be batter $b$'s $\woba$ in his $j^{th}$ plate appearance of this season, and let $\theta_b$ represent batter $b$'s unobservable ``true quality'' (the expected $\woba$ of a plate appearance with batter $b$ this season). 
After observing $j$ plate appearances, we model
%\begin{subequations}
%\label{eqn:nn}
\begin{align}
\begin{split}
\label{eqn:nn}
   x_{b1},..., x_{bj} \big| \theta_b & \sim \mathcal{N}(\theta_b, \tau^2) \\ %, \label{eqn:nn1} \\
  \theta_b & \sim \mathcal{N}(\theta_{b0}, \nu^2). %\label{eqn:nn2}
  \end{split}
\end{align}
%\end{subequations}

Here, $\theta_{b0}$ represents batter $b$'s prior ``true quality.'' 
For non-rookies, we set $\theta_{b0}$ as the average $\woba$ of a plate appearance with batter $b$ from his most recent previous season, and for rookies, we use the median $\theta_{b'0}$ over all other non-rookie batters $b'$. 
Additionally, $\nu$ represents the season-by-season standard deviation in a batter's average plate-appearance $\woba$, and $\tau$ represents the within-season standard deviation of the $\woba$ of a batter's plate appearances. 

Then, to measure batter $b$'s quality through $j$ plate appearances this season, we introduce the running-average estimator $\hat{\theta}_{bj}$ as the posterior mean $\E[\theta_b \big|  x_{b1},..., x_{bj}]$ of $\theta_b$, which as a result of our normal-normal conjugate model (\ref{eqn:nn}) is given by
\begin{equation}
\hat{\theta}_{bj} = \frac{\tau^{-2}\sum_{i=1}^{j}  x_{bi} + \nu^{-2}\theta_{b0} }{j\tau^{-2} +\nu^{-2}}.
% x^{(b)}_j = \frac{\frac{1}{\tau^2} \sum_{i=1}^{j-1}  x_{bi} + \frac{1}{\nu^2} \overline{ x_{b0}}}{\frac{j-1}{\tau^2} + \frac{1}{\nu^2}}.
\label{eqn:xtildeB}
\end{equation}
We then set $x_{j}^{(b)} = \text{logit}(\hat{\theta}_{bj}).$
% Recall that $x_{i}^{(b)}$ and $x_{i}^{(p)}$ are the logit-transformed estimates of batter and pitcher quality.
We use the logit-transformed estimates of batter quality because we
% We
felt it was more natural to allow the log-odds of each plate appearance outcome to evolve non-linearly with respect to these quality metrics.
Specifically, we find it plausible that there are diminishing returns at both extremes of player quality.
That is, we did not expect a small change in pitcher quality to manifest the same changes in the log-odds of a particular plate appearance outcome for a mediocre pitcher, an average pitcher, or an elite pitcher (keeping all else constant).
The logit transformation allows us to capture this phenomenon.
While this choice may appear somewhat unusual, we have found that it also yields a model with better predictive accuracy than a model that uses the raw quality covariates (see Appendix~\ref{sec:model_performance}).

We similarly construct a running estimate of pitcher $p$'s quality through $j$ plate appearances of the season with an analogous normal-normal model. 
For simplicity, we used the same values of $\nu$ and $\tau$ for batters and pitchers.
To set $\nu,$ we first compute the event $\woba$ for each player-season from 2006 to 2019.
Then we compute the standard deviation of these seasonal averages for each player.
The median of these player-specific standard deviations was 0.0396 for pitchers and 0.0586 for batters.
We finally set $\nu = 0.05$ to be the average of these values.
To set $\tau,$ we compute the standard deviation of event $\woba$ for each player-season from 2006 and 2016.
Across player-seasons, the median of these standard deviations was 0.509 for pitchers and 0.489 for batters.
We set $\tau = 0.5$ as a simple compromise between these values.

% %%%%% REFERENCES
% \clearpage
% \bibliography{../refs}
% \bibliographystyle{plainnat}

% \section{Simulation studies}
% \label{sec:model_sim}
% \input{model_simulations}

\section{Results}
\label{sec:results}

We fit our model to the data from each season in our dataset.
In this section, we discuss our modeling results for the 2017 season.
% In this section, we discuss our modeling results, using the 2017 season as our primary example. 
We observe qualitatively similar results in each other season. %In this section, we analyze the posterior distribution of our model fit on the observed data and discuss implications for the $\ttop$.

To obtain our posterior samples, we run four MCMC chains for 1,500 iterations.
After discarding the first 750 iterations of each chain as ``burn-in'', the Gelman-Rubin $\hat{R}$ statistic is less than $1.1$, suggesting convergence \citep{GelmanRubin1992}.
Additionally, the effective sample size of each parameter exceeds $1,172$ and the average effective sample size across all parameters is $2,852$.
It took about eight hours to run each chain.

We begin in Section~\ref{sec:little_evidence_BL} by examining the marginal posterior distributions of $\beta_{2k}$ and $\beta_{3k} - \beta_{2k},$ which quantify discontinuity in pitcher performance between successive times through the order.
As noted in Section~\ref{sec:model}, large $\ttoptwo$ or $\ttopthree$ would correspond to large, positive values of $\beta_{2k}$ or $\beta_{3k} - \beta_{2k}.$
We find, however, that the posterior distributions of these parameters are not tightly concentrated on positive values.
Instead, we find that these distributions are, for the most part, centered near zero and place substantial probability on both positive and negative values.
%We find that these distributions place considerable posterior probability in intervals containing zero.
We also see that fitted $\xwoba$ values increase steadily over the course of the game without discontinuity in the second or third time through the order. 
% Taken together, these findings suggest that our model finds little evidence of a strong batter learning effect.
Taken together, these findings suggest that our model finds little evidence of strong discontinuity between successive times through the order.

At first glance, our results appear to contradict the findings of \citet{theBook}. 
In Section~\ref{sec:tango_comp}, however, we discuss how the conclusions of \citet{theBook} actually fit within the framework of our model. 
We further find that pitcher and batter quality are much stronger predictors of $\xwoba$ than the within-game change in pitcher performance.
 % continuous pitcher decline or within-game batter learning. %(see Appendix~\ref{app:effect_sizes}).

% %%%%%%%%%%%%%%%%%%%%%%%%%%%%%%%%%%%%%%%%%%%%%%%
% \subsection{Little evidence of a strong batter learning effect}\label{sec:little_evidence_BL}
\subsection{Little evidence of strong discontinuity between successive times through the order}\label{sec:little_evidence_BL}

% First, we examine the posterior distributions of the parameters $\beta$ from our model (Equation~\eqref{eqn:model}) to gauge the significance of discontinuous changes in $\xwoba$. 
% Figure~\ref{fig:beta_boxplot_2017} shows a boxplot of the posterior distributions of these parameters.
% For each outcome $k \neq 1$, the posterior distributions of $\beta_{2k}$ and $\beta_{3k} - \beta_{2k}$ cover both positive and negative values, and most of them are centered around 0.
% For some outcomes, the posterior distributions are quite wide, which is compatible with a large effect in either direction.
% Nonetheless, we don't find strong evidence of significant discontinuity in the probability of each plate appearance outcome.
% To better interpret the effect sizes of these parameters, which are on the log odds scale, we translate these values to the the probability scale and the expected $\woba$ scale in Appendix~\ref{sec:prob_scale_TTO}.
% For instance, the posterior mean of $\beta_{3,1B}-\beta_{2,1B}$ is about $0.03$, which translates to an increase in the probability of a single from $\ttotwo$ to $\ttothree$ by about $0.002$, i.e. an increase of $1.6$ $\xwoba$ points.

\begin{figure}[hbt!]
\centering
\includegraphics[width=0.85\textwidth]{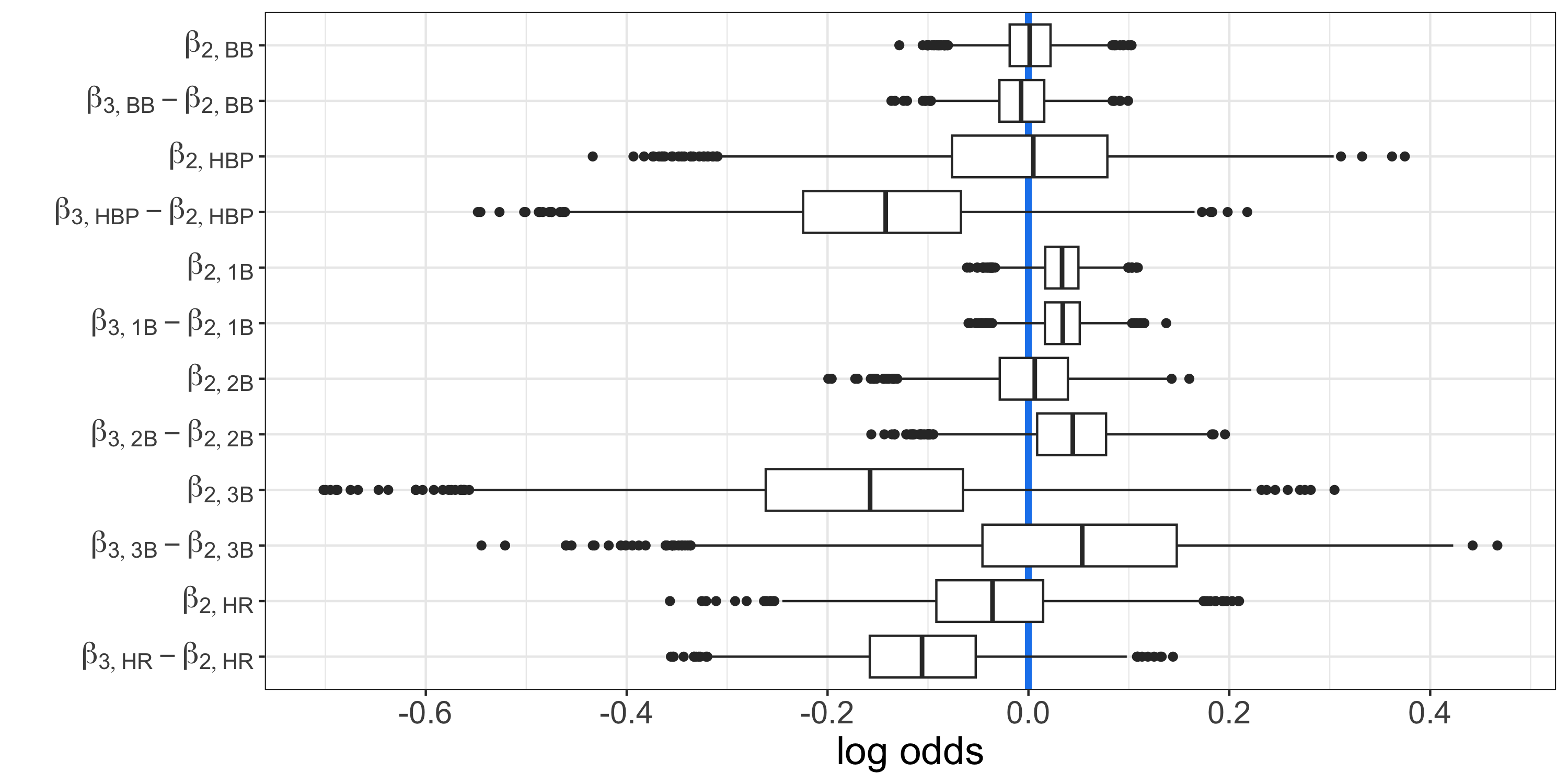}
\caption{Posterior boxplots of the $\ttop$ discontinuity parameters from Model~\eqref{eqn:model}, fit on data from 2017. The blue line denotes 0. We see that each posterior distribution covers both positive and negative values.} 
\label{fig:beta_boxplot_2017}
\end{figure}

% First, we examine the posterior distributions of the parameters $\beta$ from our model (Equation~\eqref{eqn:model}) to gauge the significance of discontinuous changes in $\xwoba$. 
First, we examine the posterior distributions of the parameters $\beta$ from our model (Equation~\eqref{eqn:model}) which control discontinuous changes in pitcher performance. 
In Figure~\ref{fig:beta_boxplot_2017} we show boxplots of the posterior distributions of the discontinuity parameters\footnote{
    To better interpret the effect sizes of these parameters, which are on the log odds scale, we translate these values to the the probability scale and the expected $\woba$ scale in Appendix~\ref{sec:prob_scale_TTO}.
} 
$\beta_{2k}$ and $\beta_{3k} - \beta_{2k}$ from our model fit on data from 2017.
Immediately we observe that none of these posterior distributions is tightly concentrated around a large positive value, which is what we would expect in the presence of a large $\ttoptwo$ or $\ttopthree.$
Instead, most of these place considerable probability on both positive and negative values.
The only exceptions are the posterior distributions of $\beta_{2,1B}$ and $\beta_{3,1B} - \beta_{2,1B},$ which measure the discontinuities in the log-odds of a single between times through the order.
Although they both place over 80\% posterior probability on the positive axis, these distributions are supported on relatively small values.
For instance, the posterior mean of $\beta_{3,1B}-\beta_{2,1B}$ is about $0.03$ on the log-odds scale, which corresponds to a change in probability no greater than 0.75 percentage points.
We additionally observe the posterior distributions corresponding to some outcomes like triples and hit-by-pitches are much more diffuse than those corresponding to other outcomes like walks and singles.
This is not entirely unexpected: there are considerably more singles and walks in the dataset than triples and hit-by-pitches and the relative uncertainties about the corresponding $\beta_{2k}$ and $\beta_{3k} - \beta_{2k}$ values closely track the frequencies of these outcomes. 
% Ultimately, we do not find the posterior distributions in Figure~\ref{fig:beta_boxplot_2017} to be consistent with large, systematic time through the order penalties.
Ultimately, we do not find the posterior distributions in Figure~\ref{fig:beta_boxplot_2017} to be indicative of large, systematic time through the order penalties.
We obtain similar findings in each season from 2012 to 2019 (see Figure~\ref{fig:beta_boxplot_ALLYRS} in Appendix~\ref{sec:trend_across_years}).
%For some outcomes (e.g. walks and singles), the posterior distributions are tightly concentrated around 0, and for other outcomes (e.g., triples and hit-by-pitches, which are rare events), the posterior distributions are quite wide, which is compatible with a large effect in either direction.
%Overall, the posterior distributions of the discontinuity parameters cover both positive and negative values, and most of them are centered around 0.
%In particular, we don't see what we would expect to see if there were strong evidence for a $\ttop$ -- i.e., we don't see the posterior distributions tighly concentrated around a positive number.
%Thus we don't find overwhelming evidence of strong discontinuity in the probability of each plate appearance outcome.
% Rather, given our model and one season of data, there is considerable uncertainty over the existence of a $\ttop$, and our boxplots are not inconsistent with no effect.
%Moreover, in Figure~\ref{fig:beta_boxplot_ALLYRS} of Appendix~\ref{sec:trend_across_years} we see a similar trend in the posterior boxplots from each year from 2012 to 2019.
%Thus, we do not find overwhelming evidence for a consistent $\ttop$ across the last decade.

%%%%%%%%%%%%%%%%%%%
%%%%%%%%%%%%%%%%%%%

\begin{figure}[hbt!]
\centering
\includegraphics[width=0.7\textwidth]{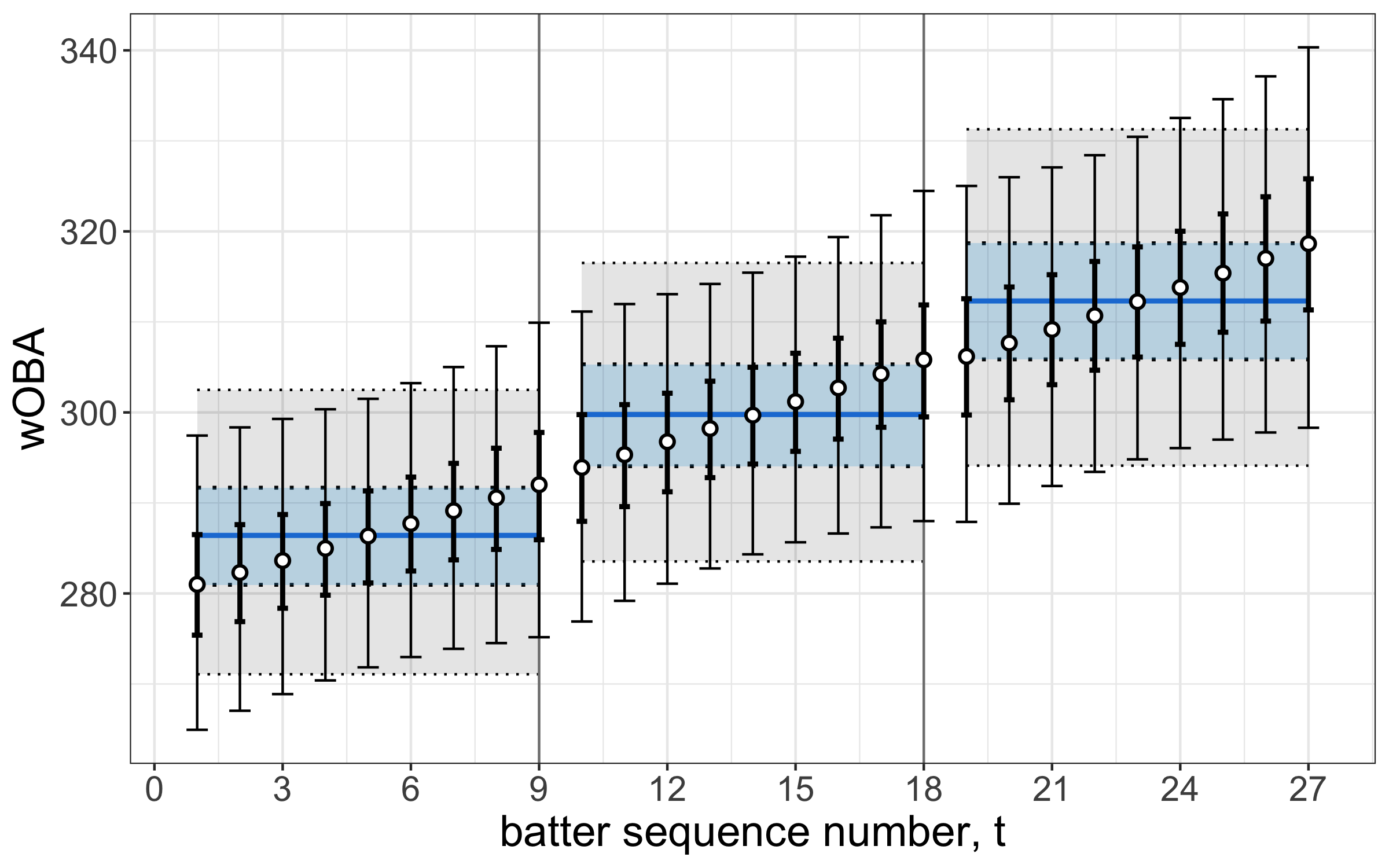}
\caption{
Trend in expected $\woba$ over the course of a game in 2017 for an average batter facing an average pitcher of the same handedness on the road.
The white dots, thick black bars, and thin black bars denote the posterior mean, 50\% credible interval, and 95\% credible interval of $\xwoba(t,\tilde{\bx}).$ The blue lines, blue shaded regions, and gray shaded regions denote the posterior mean, 50\% credible interval, and 95\% credible interval of $\xwoba(t,\tilde{\bx})$ averaged within each $\tto$.
} 
% \label{fig:plot_xwoba_over_time_shading}
\label{fig:plot_xwoba_over_time_2017a}
\end{figure}

Furthermore, we plot the trajectory of a pitcher's expected $\woba$ over the course of the game according to our model, fit on data from 2017.
%, for a batter of average quality on the road facing a pitcher of average quality with a handedness match. 
Specifically, in Figure~\ref{fig:plot_xwoba_over_time_2017a}, we plot the posterior distribution of the sequence of $\xwoba(t,\tilde{\bx}),$ where $\tilde{\bx}$ corresponds to an average batter facing an average pitcher of the same handedness on the road, %(see Equation~\eqref{eqn:x_tilde}).
\begin{equation}
\label{eqn:x_tilde}
% \tilde{\bx}^\top = (\text{logit}(0.315), \text{logit}(0.315), 1, 0).
\tilde{\bx}^\top = (\overline{x^{(b)}}, \overline{x^{(p)}}, 1, 0).
\end{equation}
The white dots, thick black bars, and thin black bars denote the posterior mean, 50\% credible interval, and 95\% credible interval of $\xwoba(t,\tilde{\bx}).$
For now, ignore the blue lines, blue shaded regions, and gray shaded regions, which we explain the next Section~\ref{sec:tango_comp}.
We see that expected $\woba$ increases steadily over the course of a game, without discontinuity 
% due to batter learning 
in the second or third time through the order. 
In other words, our model finds little evidence for a strong discontinuity in the expected $\woba$ of a plate appearance.
% In other words, our model finds little evidence for an effect of batter learning on the expected $\woba$ of a plate appearance.
This trend is persistent across each year from 2012 to 2019 (see Figure~\ref{fig:xwoba_trend_across_years} of Appendix~\ref{sec:trend_across_years}) and other choices of $\bx$.

% %%%%%%%%%%%%%%%
% \begin{figure}[H]
% \centering
% \includegraphics[width=0.7\textwidth]{plots/plot_obs_results_2017_xwoba_check.png}
% \caption{
% Trend in expected $\woba$ over the course of a game in 2017 for an average batter facing an average pitcher of the same handedness on the road.
% The white dots indicate the posterior means of the expected $\woba$ values, the thick black error bars denote the 50\% credible intervals, and the thin black error bars denote the 95\% credible intervals.} 
% \label{fig:plot_xwoba_over_time_2017a}
% \end{figure}
% %%%%%%%%%%%%%%%

%%%%%%%%%%%%%%%%%
%%%%%%%%%%%%%%%

%%%%%%%%%%%%%%%%%%%%%%%%%%%%%%%%%%%%%%%%%%%%%%%
\subsection{\citet{theBook}'s conclusions fit within our framework}\label{sec:tango_comp}

At first glance, our results appear to contradict the findings of \citet{theBook}.
Recall, however, that while we carefully estimate the $\xwoba$ for each batter faced, \citet{theBook} identified the $\ttop$ by comparing $\woba$ averaged across entire times through the order.
By similarly averaging $\xwoba(t,\bx)$ within times through the order, it turns out that we can recover the $\ttop$ identified by \citet{theBook}.

Formally, for each plate-appearance-state vector $\bx,$ consider the average difference of $\xwoba(t,\bx)$ between the first and second times through the order,
\begin{equation}
\label{eqn:D12x}
\mathscr{D}_{12}(\bx) = \frac{1}{9} \sum_{t=1}^{9} \left[ \xwoba(t + 9,\bx) - \xwoba(t,\bx) \right].
% \mathscr{D}_{12x} = \frac{1}{9} \sum_{t=1}^{9} \mathscr{D}_{12tx},
\end{equation}
Using our fitted model, we study the posterior distribution of $\mathscr{D}_{12}(\bx)$ and the similarly defined $\mathscr{D}_{23}(\bx),$ which captures the change in average $\xwoba$ between the second and third $\tto$.

The posterior means of $\mathscr{D}_{12}(\tilde{\bx})$ and $\mathscr{D}_{23}(\tilde{\bx})$ are about 13 $\woba$ points, which are consistent with \citet{theBook}'s findings. 
Also, virtually all of the posterior samples are positive, suggesting that average pitcher performance indeed declines from one $\tto$ to the next. 
Specifically, our model suggests that the expected $\woba$ points of an average plate appearance increases by $13.4$ (with a $95\%$ credible interval of [7.78, 19.0]) from the first $\tto$ to the second, and by $12.5$ (with a $95\%$ credible interval of [5.98, 18.7]) from the second $\tto$ to the third. 
We show histograms of the posterior samples of $\mathscr{D}_{12}(\tilde{\bx})$ and $\mathscr{D}_{23}(\tilde{\bx})$ in Figure~\ref{fig:tto_diff_xw} in Appendix~\ref{sec:prob_scale_TTO}.

% Like Figure~\ref{fig:plot_xwoba_over_time_2017a}, Figure~\ref{fig:plot_xwoba_over_time_2017a} 
Figure~\ref{fig:plot_xwoba_over_time_2017a} overlays the trajectory of $\xwoba(t,\tilde{\bx})$ with the posterior mean (the blue lines), the 50\% credible intervals (the blue shaded regions), and the 95\% posterior credible intervals (the gray shaded regions) of the $\xwoba(t,\tilde{\bx})$ trajectory  averaged over each $\tto$. 
We see that mean pitcher performance within a $\tto$ declines from each $\tto$ to the next by about 13 $\woba$ points.
Figure~\ref{fig:plot_xwoba_over_time_2017a} reveals how these declines in average performance are an 
% artifact of continuous pitcher decline rather than discontinuous batter learning.
artifact of continuous, not discontinuous, pitcher decline.

%%%%%%%%%%%%%%%%%%%%%%%%%%%%%%%%%%%%%%%%%%%%%%%%%%%%%%%%%%%%%%%%%%%%%%%%%%%%%%%%%%%%%%%%%
\subsection{The impact of handedness match and home field advantage on the outcome of a plate appearance}

As discussed previously, pitchers decline from one TTO to the next by about 13 $\woba$ points on average. 
Now, we compare this effect size to that of confounders like batter quality, pitcher quality, handedness match, and home field advantage. 
We find that batter quality and pitcher quality have a much larger impact on predicting the outcome of a plate appearance, whereas handedness and home field advantage have a similar effect size as the batter sequence number. 

We begin by assessing the impact of handedness match and home field advantage on the outcome of a plate appearance. 
To do so, we compute the posterior mean of the expected $\woba$ of a plate appearance averaged over the batter sequence numbers, for different combinations of handedness and home field advantage. 
Mathematically, for a batter of average quality with batter-at-home value $\texttt{home} \in \{0,1\}$ facing a pitcher of average quality having handedness match value $\texttt{hand} \in \{0,1\}$, yielding plate-appearance-state vector 
\begin{equation}
\label{eqn:pa_sv_hh}
% \bx^\top = (\text{logit}(0.315), \text{logit}(0.315), \texttt{home} , \texttt{hand} ),
\bx^\top = (\overline{x^{(b)}}, \overline{x^{(p)}}, \texttt{home} , \texttt{hand} ),
\end{equation}
% we compute
% \begin{equation}
% \label{eqn:hand_home_xwoba_comp_Formula}
% \E \bigg[ \frac{1}{27} \sum_{t=1}^{27} \xwoba(t,\bx) \bigg].
% \end{equation}
we compute the posterior mean and standard deviation of 
\begin{equation}
\label{eqn:hand_home_xwoba_comp_Formula}
\frac{1}{27} \sum_{t=1}^{27} \xwoba(t,\bx).
\end{equation}

% In Table~\ref{table:hand_home_combos} we show the posterior mean, and in parenthesis twice the posterior standard deviation, of Formula~\eqref{eqn:hand_home_xwoba_comp_Formula} for all combinations of $\texttt{hand}$ and $\texttt{home}$. 
In Table~\ref{table:hand_home_combos} we show the posterior mean $\pm$ two posterior standard deviations\footnote{
    Although the posterior distribution of Formula~\eqref{eqn:hand_home_xwoba_comp_Formula} is not exactly Gaussian, we find that the actual $95\%$ credible interval is extremely close to the interval computed as the posterior mean $\pm$ twice the standard deviation.
} 
of Formula~\eqref{eqn:hand_home_xwoba_comp_Formula} for all combinations of $\texttt{hand}$ and $\texttt{home}$. 
Home field advantage has a similar effect size as pitcher decline across one $\tto$:  a batter at home has about 12 more mean expected $\woba$ points than a batter on the road. 
Handedness match has a slightly larger effect:  a pitcher whose handedness matches that of the batter has about 18 fewer mean expected $\woba$ points than one whose handedness does not match. 
The $\xwoba$ intervals, given by the posterior mean $\pm$ two posterior standard deviations, overlap for a batter at home vs. away but do not overlap for a batter with vs. without a handedness match.
In other words, we find a significant handedness effect but not a significant home field effect.

\begin{table}[H]
\centering
\begin{tabular}{ cc|cc } \hline
&  & \multicolumn{2}{c}{Batter at Home}  \\
&  & $0$ & $1$  \\ \hline 
\multirow{2}{*}{\rotatebox[origin=c]{90}{Hand}}
\multirow{2}{*}{\rotatebox[origin=c]{90}{Match}}
& $0$ & $316$ ($\pm 7.8$) & $328$ ($\pm 7.8$)  \\[0.2cm]
& $1$ & $298$ ($\pm 6.9$) & $310$ ($\pm 7.2$)  \\
\hline
\end{tabular}
\caption{For different combinations of handedness match and home field advantage, the posterior mean (and, in parenthesis, twice the posterior standard deviation) of the expected $\woba$ points of a plate appearance, assuming a batter of average quality faces a pitcher of average quality, averaged over the batter sequence numbers $t=1,...,27$.}
\label{table:hand_home_combos}
\end{table}

%%%%%%%%%%%%%%%%%%%%%%%%%%%%%%%%%%%%%%%%%%%%%%%%%%%%%%%%%%%%%%%%%%%%%%%%%%%%%%%%%%%%%%%%%
\subsection{The impact of batter quality and pitcher quality on the outcome of a plate appearance}\label{sec:bq_pq}

Now, we assess the impact of batter quality and pitcher quality on the outcome of a plate appearance.
To do so, for different combinations of batter and pitcher quality, we compute the posterior mean of the expected $\woba$ of a plate appearance, averaged over the batter sequence numbers $t \in \{1,...,27\}$. Mathematically, for a batter of quality $x^{(b)}$ on the road facing a pitcher of quality $x^{(p)}$ with a handedness match, yielding plate-appearance-state vector 
\begin{equation}
\label{eqn:pa_sv_bqpq}
\bx^\top = (x^{(b)}, x^{(p)}, 1, 0),
\end{equation}
we compute the posterior mean and standard deviation of
\begin{equation}
\label{eqn:bq_pq_xwoba_comp_Formula}
\frac{1}{27} \sum_{t=1}^{27} \xwoba(t,\bx).
% \E \bigg[ \frac{1}{27} \sum_{t=1}^{27} \xwoba(t,\bx) \bigg].
\end{equation}

% In Table \ref{table:bq_pq_combos} we show the posterior mean, and in parenthesis twice the posterior standard deviation, of Formula~\ref{eqn:bq_pq_xwoba_comp_Formula} for all combinations of the $25^{th}, 50^{th}, \text{ and } 75^{th}$ quantiles of $x^{(b)}$ and $x^{(p)}$. 
In Table \ref{table:bq_pq_combos} we show the posterior mean $\pm$ two posterior standard deviations of Formula~\ref{eqn:bq_pq_xwoba_comp_Formula} for all combinations of the $25^{th}, 50^{th}, \text{ and } 75^{th}$ quantiles of $x^{(b)}$ and $x^{(p)}$. 
Specifically, we take the quantiles of the empirical distributions from Figure \ref{fig:woba_dists} from Section \ref{sec:wOBA}. 
For batters, the $25^{th}$ quantile represents a bad batter, the $50^{th}$ an average batter, and the $75^{th}$ a good batter.
Conversely, for pitchers, 
the $25^{th}$ quantile represents a good pitcher, the $50^{th}$ an average pitcher, and the $75^{th}$ a bad pitcher.

As shown in Table \ref{table:bq_pq_combos}, the quality of the batter and pitcher has a larger impact on the outcome of a plate appearance than the batter sequence number $t \in \{1,...,27\}$. 
For instance, fix a batter's quality. 
The difference in mean expected $\woba$ points between a good and bad pitcher is large: about $42$ to $48$ $\woba$ points, depending on the batter quality. 
To see this, consider the second row of Table \ref{table:bq_pq_combos}, in which a median batter ($50^{th}$ quantile) faces pitchers of various quality, assuming the batter is on the road and has the same handedness as the pitcher, averaged over each lineup position.
The expected $\woba$ points of a plate appearance against a good pitcher ($25^{th}$ quantile) is $288$, and against a bad pitcher ($75^{th}$ quantile) is $333$. 
So, for a median batter, the difference in expected $\woba$ points between a good and a bad pitcher is about $45$ $\woba$ points.

Conversely, fix a pitcher's quality. 
Then the difference in mean expected $\woba$ points between a good and bad batter is also large: about 36 to 41 $\woba$ points, depending on the pitcher quality. 
Finally, note that these effects are significant, as the intervals given by the posterior mean $\pm$ two posterior standard deviations do not overlap.

%%%%%%%%%%%%%%
\begin{table}[H]
\centering
\begin{tabular}{ cc|ccc } \hline
& &  \multicolumn{3}{c}{Pitcher Quality} \\
& & 
\text{$25^{th}$ quantile} &
\text{$50^{th}$ quantile} &
\text{$75^{th}$ quantile} \\ \hline
% \multirow{5}{*}{\rotatebox[origin=c]{90}{Batter Quality}}
\multirow{3}{*}{\rotatebox[origin=c]{90}{Batter}}
\multirow{3}{*}{\rotatebox[origin=c]{90}{Quality}}
& \text{$25^{th}$ quantile} & $270$ ($\pm 6.7$) & $291$ ($\pm 6.9$) & $313$ ($\pm 7.5$) \\
& \text{$50^{th}$ quantile} & $288$ ($\pm 7.0$) & $310$ ($\pm 7.1$) & $333$ ($\pm 7.7$)  \\
& \text{$75^{th}$ quantile} & $306$ ($\pm 7.7$) & $329$ ($\pm 7.8$) & $354$ ($\pm 8.5$)  \\
\hline
\end{tabular}
\caption{For different combinations of batter quality and pitcher quality (in terms of $\woba$ points) -- in particular, the $25^{th}, 50^{th}, \text{ and } 75^{th}$ quantile -- the posterior mean (and, in parenthesis, twice the posterior standard deviation) of the expected $\woba$ points of a plate appearance, assuming batters are on the road and have the same handedness as the pitcher, averaged over the batter sequence numbers $t=1,...,27$.}
\label{table:bq_pq_combos}
\end{table}

Therefore, pitcher quality and batter quality have a much larger impact on the outcome of a plate appearance than within-game pitcher decline.

% %%%%%%%%%%%%%%
% \begin{table}[H]
% \centering
% \begin{tabular}{ cc|ccccc } \hline
% & &  & \multicolumn{3}{c}{Pitcher Quality}  \\
% & & 
% \text{$5^{th}$ quantile} &
% \text{$25^{th}$ quantile} &
% \text{$50^{th}$ quantile} &
% \text{$75^{th}$ quantile} &
% \text{$95^{th}$ quantile}  \\ \hline
% \multirow{5}{*}{\rotatebox[origin=c]{90}{Batter Quality}}
% & \text{$5^{th}$ quantile} & 197 & 232 & 253 & 272 & 306 \\
% & \text{$25^{th}$ quantile} & 226 & 266 & 289 & 31q & 349 \\
% & \text{$50^{th}$ quantile} & 242 & 285 & 309 & 332 & 373 \\
% & \text{$75^{th}$ quantile}  & 260 & 305 & 331 & 355 & 399 \\
% & \text{$95^{th}$ quantile}  & 291 & 341 & 370 & 396 & 443 \\
% \hline
% \end{tabular}
% \caption{For different combinations of batter quality and pitcher quality (in terms of wOBA points) -- in particular, the $5^{th}, 25^{th}, 50^{th}, 75^{th},$ and $95^{th}$ quantiles -- the posterior mean of the expected wOBA points of a plate appearance, assuming batters are on the road and have the same handedness as the pitcher, averaged over the batter sequence numbers $t=1,...,27$.}
% \label{table:bq_pq_combos}
% \end{table}

% % %%%%% REFERENCES
% \clearpage
% \bibliography{../refs}
% \bibliographystyle{plainnat}

\section{Discussion}
\label{sec:baseball_implications}

It has long been observed that batters tend to perform better the more times they face a particular pitcher. 
\citet{theBook} first quantified the corresponding drop-off in pitcher quality and attributed the apparent time through the order penalty to batter learning.
% Their analysis, however, does not attempt to disentangle batter learning from pitcher fatigue.
Their analysis, however, does not attempt to disentangle continuous evolution in pitcher performance over the course of the game from discontinuities between successive times through the order.
%batter learning from pitcher fatigue.   % by design
We instead model the outcome of a plate appearance in a way that accommodates both of these.
 % batter learning and continuous pitcher decline.
Our analysis reveals the expected $\woba$ of a plate appearance increases steadily over the course of the game, over average, without significant discontinuity between each time through the order.
% Additionally, the posterior distributions of the model parameters that quantify batter learning cover both positive and negative values. 
Additionally, the posterior distributions of the model parameters that quantify discontinuous pitcher decline cover both positive and negative values. 
These results suggest there is little evidence of strong discontinuity in pitcher performance between successive times through the order.
% These results suggest there is little evidence of a strong batter learning effect.
Based on our analysis, we do not believe it always appropriate to pull pitchers at the start of the third time through the order.
Rather, we recommend managers base their decisions to pull a pitcher on a pitcher's quality and continuous decline throughout the game.

Although \citet{theBook} attributes within-game pitcher decline to batter learning, we hesitate to make conclusions about the potential causes of within-game pitcher decline.
Nonetheless, we offer potential interpretations of the parameters of our model from Equation~\eqref{eqn:model}.
Because a batter faces the opposing team's pitcher at most once in each $\tto$, it is natural to interpret the parameters $\beta_{2k}$ and $\beta_{3k} - \beta_{2k}$ which quantify discontinuous pitcher evolution as batter learning parameters.
A pitcher, on the other hand, faces each opposing batter.
Thus it is natural to interpret the parameters $\alpha_{0k}$ and $\alpha_{1k}$ which quantify continuous pitcher decline as pitcher fatigue parameters. 
In particular, it is known that pitchers fatigue continuously over the course of a game (e.g., \citet{greenhouse13}).
Nonetheless, there are other potential mechanisms of pitcher decline (e.g., a changing pitch selection, discussed below), and we don't explicitly adjust for pitcher fatigue.
% , or even measure or quantify pitcher fatigue at all.
% Hence we hesitate to make conclusions about the causes behind our finding that there is little evidence of strong discontinuity in pitcher performance between sucessive times through the order.
Hence we hesitate to make causal conclusions from our model.

% \ryan{
% interpretations of batter learning and pitcher fatigue. \\
% It has long been observed that batters tend to perform better the more times they face a particular pitcher. 
% \citet{theBook} first quantified the corresponding drop-off in pitcher quality and attributed the apparent time through the order penalty to batter learning.
% }

% Although our analysis is more nuanced than \citet{theBook}'s, our analysis is not without limitations.
% Recall that our model allows the log-odds of each non-out plate appear outcome to evolve linearly with batter sequence number.
% To make our model more flexible, one could replace the $\alpha_{0k} + \alpha_{1k}t$ terms in Equation~\eqref{eqn:model} with a spline expansion.
% It is plausible that pitcher performance declines at different rates.
% To account for such heterogeneity, we could extend our model by introducing pitcher-specific rates of decline rather than using a common $\alpha_{1k}$ term. 
% Such extension would yield a model similar to that in Equation 1 of \citet{age_curves} and could be fitted with a hierarchical Bayesian model.
% We additionally note that we hold our measures of pitcher and batter quality fixed throughout a single game.
% While it is certainly possible to update our running-average estimates of player quality after each plate appearance, we note that there is considerably more variation between-games than within-game.

% Although
Furthermore, although our analysis is more nuanced than \citet{theBook}'s, our analysis is not without limitations.
Recall that our model allows the log-odds of each non-out plate appearance outcome to evolve linearly with batter sequence number.
A more flexible model wouldn't force a particular functional form on the change in pitcher performance over the course of a game.
We find that using a more flexible model doesn't change the qualitative results of our study (see Appendix~\ref{sec:indicator_model}).
Additionally, our model assumes that the trajectory of within-game pitcher deterioration is the same across all pitchers and batters.
A more elaborate model would allow within-game performance to change at at different rates for different players.
We find that using this more elaborate model doesn't change the qualitative results of our study (see Appendix~\ref{sec:varying_slopes_model}). 
% Additionally, we note that there is enormous variation in pitching performance on a game-by-game basis.
% Although \citet[Chapter 7]{theBook} believe this is due to randomness rather than pitcher ``hotness'', a more flexible model may use an estimate of pitcher quality which updates as a game evolves.

\begin{figure}[h]
\centering
\includegraphics[width = \textwidth]{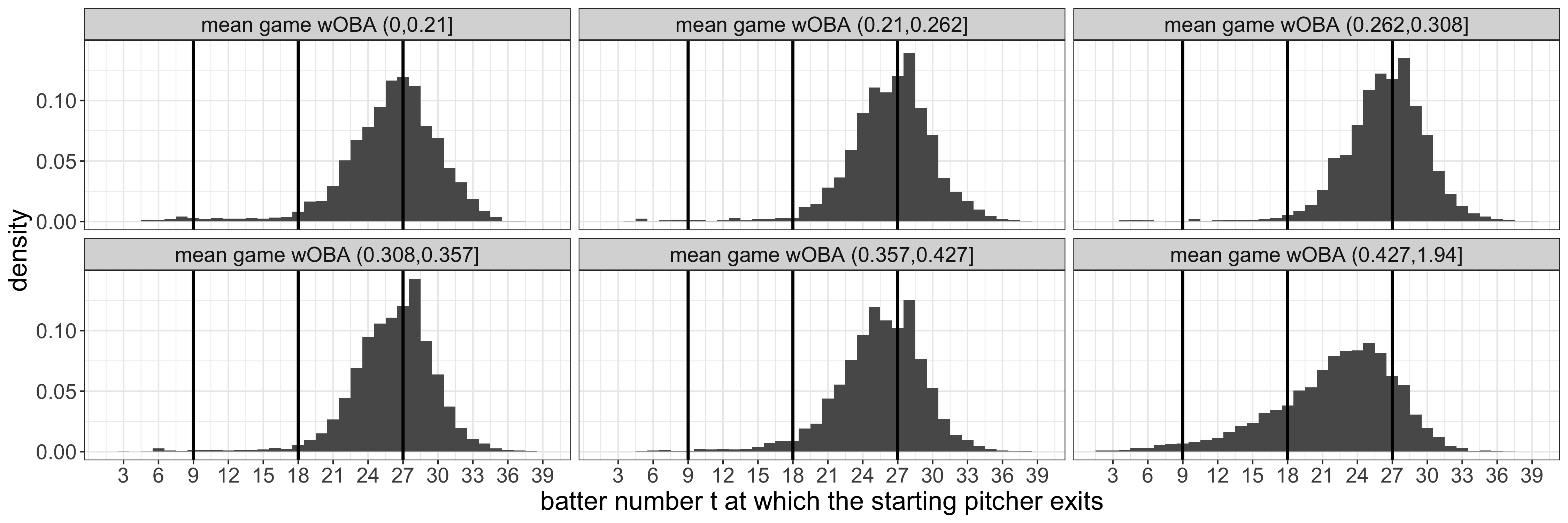}
\caption{Histogram of the batter sequence number $t$ at which a starting pitcher exits the game, 
for different bins of mean game $\woba$.
% for six quantile bins of mean game $\woba$.
}%
\label{fig:visualize_selection_bias2}
\end{figure}

Additionally, we note that there is enormous variation in pitching performance on a game-by-game basis.
Although \citet[Chapter 7]{theBook} believe this is due to randomness rather than pitcher ``hotness'', a more flexible model may use an estimate of pitcher quality which updates as a game evolves.
For those who believe in pitcher ``hotness'', omitting a measure of within-game pitcher quality contributes further to selection bias.
In particular, whether we observe a pitcher in $\ttothree$ depends on his performance earlier in the game, as a pitcher who ``bombs'' or begins pitching poorly is more likely to be removed earlier in the game.
We visualize this survival process in Figure~\ref{fig:visualize_selection_bias2}, which shows that pitchers who have a bad pitching day (mean game $\woba$ larger than, say, $0.437$) are much more likely to be removed earlier in the game.
Note that the six mean game $\woba$ bins in Figure~\ref{fig:visualize_selection_bias2} are six evenly sized quantiles of mean game $\woba$.
% The mean game $\woba$ bins in Figure~\ref{fig:visualize_selection_bias2} are six quantiles of 
So, a starting pitcher who remains in $\ttothree$ pitched better that day over average than one who is pulled prior to $\ttothree$, and it is plausible that the former pitcher would be better in $\ttothree$ than the latter pitcher.
On this view, our approach underestimates the magnitude of continuous pitcher decline.
But, as discussed in Section~\ref{sec:selection_bias_adjustment}, our goal is to estimate the discontinuous decline parameters $\beta$, which our approach does a reasonable job of; we leave a more elaborate estimation of continuous pitcher decline to future work.

%%%%%%%%%%%%%%%%%%%%%
%%%%%%%%%%%%%%%%%%%%%

Furthermore,
our analysis does not account for pitch selection, which, for some pitchers, evolves over the course of the game.
Changes in pitch selection may be a response to pitcher fatigue: for instance, the more tired a pitcher becomes, the more difficult it may be to throw a fastball.
Alternatively, pitchers might change their pitches in response to perceived batter learning: to prevent batters from learning his tendencies, a pitcher can perhaps be more unpredictable by changing his pitch selection over the game.
A more fine-grained analysis would capture this within-game change in pitch selection, perhaps by modeling pitcher quality as a function of pitch selection.
%Nonetheless, a change in pitch selection over the course of the game may also present itself as a continuous process.
% On this view, 
Nonetheless, modeling a pitcher's continuous change over the game may simultaneously adjust for pitcher fatigue and an evolving pitch selection.

Additionally, recall that we use an empirical Bayes approach to quantify batter and pitcher quality.
Specifically, early in the season we let a player's quality be close to his average $\woba$ from the end of his previous season, and later in the season be closer to his current average $\woba$.
Our current analysis shrinks to the prior season's average $\woba$ similarly for all players (e.g., the prior variance is constant).
But, the more we've observed a player in the past, the more confident we should be in the player's ability this season.
Thus a more fine-grained analysis would employ a more flexible empirical Bayes approach which allows the prior variance to vary in the number of last season's plate appearances (e.g., see \citet{Brown2008}).
Additionally, a more elaborate approach may shrink to some combination of a pitcher's previous season mean $\woba$ and the overall mean of pitcher quality from the previous season, rather than shrinking to just the former.

\section*{Acknowledgements}
% Acknowledgements

The authors thank Tom Tango for his comments on an early draft of this paper.
The authors acknowledge the High Performance Computing Center (HPCC) at The Wharton School, University of Pennsylvania for providing computational resources that have contributed to the research results reported within this paper.

% Funding
Support for S.K.D. was provided by the University of Wisconsin--Madison, Office of the Vice Chancellor for Research and Graduate Education with funding from the Wisconsin Alumni Research Foundation.

%\FloatBarrier
\bibliography{refs}

\begin{thebibliography}{}

\bibitem[Brown, 2008]{Brown2008}
Brown, L.~D. (2008).
\newblock {In-season prediction of batting averages: A field test of empirical
  Bayes and Bayes methodologies}.
\newblock {\em The Annals of Applied Statistics}, 2(1):113 -- 152.

\bibitem[Carpenter et~al., 2017]{Stan}
Carpenter, B., Gelman, A., Hoffman, M.~D., Lee, D., Goodrich, B., Betancourt,
  M., Brubaker, M., Guo, J., Li, P., and Riddell, A. (2017).
\newblock {Stan: A Probabilistic Programming Language}.
\newblock {\em Journal of Statistical Software}, 76(1):1--32.

\bibitem[Fangraphs, 2021]{wobaCharts}
Fangraphs (2021).
\newblock {wOBA and FIP Constants}.
\newblock \\ \url{https://www.fangraphs.com/guts.aspx?type=cn}.

\bibitem[Gelman and Rubin, 1992]{GelmanRubin1992}
Gelman, A. and Rubin, D.~B. (1992).
\newblock Inference from iterative simulation using multiple sequences.
\newblock {\em Statistical Science}, 7:457--472.

\bibitem[Greenhouse, 2011]{greenhouse13}
Greenhouse, J. (2011).
\newblock {Spitballing: Fourth Time's the Harm}.
\newblock \\
  \url{https://www.baseballprospectus.com/news/article/13117/spitballing-fourth-times-the-harm/}.

\bibitem[Laurila, 2015]{MLBmanagersTTOPquotes}
Laurila, D. (2015).
\newblock {Managers on the Third Time Through the Order}.
\newblock \\
  \url{https://blogs.fangraphs.com/managers-on-the-third-time-through-the-order/}.

\bibitem[Lichtman, 2013]{bballProspectusTTO}
Lichtman, M. (2013).
\newblock {Baseball ProGUESTus: Everything You Always Wanted to Know About the
  Times Through the Order Penalty}.
\newblock \\ \url{https://www.baseballprospectus.com/news/article/22156/}.

\bibitem[{R Core Team}, 2020]{R}
{R Core Team} (2020).
\newblock {\em R: A Language and Environment for Statistical Computing}.
\newblock R Foundation for Statistical Computing, Vienna, Austria.

\bibitem[Rivera, 2020]{rivera20}
Rivera, J. (2020).
\newblock {Rays' Kevin Cash explains decision to pull Blake Snell in World
  Series: 'I regret it because it didn't work out'}.
\newblock \\
  \url{https://www.sportingnews.com/us/mlb/news/kevin-cash-blake-snell-world}
  \\ \url{-series-explained/lfnyfc4nqwys1pcncc2lnyjho}.

\bibitem[Slowinski, 2010]{wOBAExplanation}
Slowinski, P. (2010).
\newblock {wOBA}.
\newblock \\ \url{https://library.fangraphs.com/offense/woba/}.

\bibitem[{Stan Development Team}, 2022]{rstan}
{Stan Development Team} (2022).
\newblock {\em {RStan}: the {R} interaface for {Stan}}.

\bibitem[Tango et~al., 2007]{theBook}
Tango, T., Lichtman, M., and Dolphin, A. (2007).
\newblock {\em {The Book: Playing the Percentages in Baseball}}.
\newblock Potomac Books.

\end{thebibliography}

%% prepare supplementary stuff here
%\renewcommand{\theequation}{S\arabic{equation}}
%\renewcommand{\thesection}{S\arabic{section}}  
%\renewcommand{\thefigure}{S\arabic{figure}}  
%\renewcommand{\thetable}{S\arabic{table}} 
%\renewcommand{\themyLemma}{S\arabic{myLemma}} 
%\renewcommand{\themyTheorem}{S\arabic{myTheorem}} 
%\renewcommand{\themyProposition}{S\arabic{myProposition}}
%%\renewcommand{\thesubsection}{S\arabic{subsection}} 
%%\renewcommand{\thesubsubsection}{S\arabic{subsubsection}} 
%\setcounter{equation}{0}
%\setcounter{section}{0}
%\setcounter{subsection}{0}
%\setcounter{subsubsection}{0}
%\setcounter{myLemma}{0}
%\setcounter{myTheorem}{0}

%\newpage
%\begin{center}
%{\LARGE \textbf{Supplementary Materials}}
%\end{center}
\appendix
% %%%%%%%%%%%%%%%%%%%%%%%%%%%%%
% \usepackage{fullpage, parskip}
% \onehalfspacing
% %%%%%%%%%%%%%%%%%%%%%%%%%%%%%

%%%%%%%%%%%%%%%%%%%%%%%%%%%%%%%%%%%%%%%%%%%%%%%%%%%%%%%%%%%%%%%
%%%%%%%%%%%%%%%%%%%%%%%%%%%%%%%%%%%%%%%%%%%%%%%%%%%%%%%%%%%%%%%

%%%%%%%%%%%%%%%%%%%%%%%%%%%%%%%%%%%%%%%%%%%%%%%%%%%%%%%%%%%%%%%
%%%%%%%%%%%%%%%%%%%%%%%%%%%%%%%%%%%%%%%%%%%%%%%%%%%%%%%%%%%%%%%
\section{Our code and data}

Our code is available on Github\footnote{\url{https://github.com/snoopryan123/TTO_}}. The \texttt{data\_wrangling} folder of the Github repository contains our dataset processing, including the Retrosheet data scraper. The \texttt{data} folder further processes the full dataset into a smaller dataset relevant for this paper. Finally, the \texttt{model\_positive\_slope\_prior} folder contains our data analysis, including our \texttt{Stan} model. 

The final datasets used in this paper are available for download\footnote{\url{https://upenn.app.box.com/folder/144635702840?v=retrosheet-pa-1990-2000}}. The cleaned dataset of all MLB plate appearances from 1990 to 2020 is \texttt{retro\_final\_PA\_1990-2020d.csv}. The datasets $\texttt{TTO\_dataset\_410.csv}$ and $\texttt{TTO\_dataset\_510.csv}$ are processed subsets of the large dataset which we use to fit our models.

\section{Model simulation study}\label{sec:model_sim}

We conduct a simulation study to assess the capacity of our model (Equation~\eqref{eqn:model}) to estimate time through the order penalties of various sizes.
Specifically, we simulate data consistent with different $\ttops$ and verify that our posterior estimates are close to the data generating parameters. 

\textbf{Simulation setup.}
For our first simulation, we generate data consistent with continuous pitcher fatigue and no $\ttop$ for any of the plate appearance outcomes by setting $\beta_{2k} = \beta_{3k} = 0$ for each $k \neq 1.$
In our second simulation, for each $k \neq 1,$ we set the $\beta_{2k}$ and $\beta_{3k}$ so that the resulting $\xwoba$ curves display $\ttops$ consistent with \citet{theBook}'s findings of about 10 expected $\woba$ points between successive times through the order.
%the alleged effect size from \citet{theBook} of about 10 $\woba$ points between each time through the order.
Finally, for our third simulation, we set $\beta_{2k}$ and $\beta_{3k}$ so that there is no $\ttoptwo$ (in terms of $\xwoba$) but a large $\ttopthree$ of about 50 $\woba$ points. 
For each simulation, we set the values of the $\alpha_{0k}$'s, $\alpha_{1k}$'s, and $\eta_{k}$'s in a way that is consistent with observed data. 
Additional details about the simulation setup, including the data generating parameter values, are available in Appendix~\ref{app:simulation_details}.

For each simulation, we generate 225 full seasons worth of data.
We fit our model to 80\% of the data from each simulated season and evaluate our fitted model's predictive performance on the remaining 20\%.
We further assess how well our fitted model recovers the function $\xwoba(t,\bx)$ for a set of average confounder values. 
% In particular, \skd{explain this bit further}

%We then gather our observed data $\{t_i\}$ and $\{X_{i\ast}\}$ from 2017, consisting of every plate appearance in 2017. 
%Then, for each of our three simulation studies, we use our ``true'' simulation parameters to generate 25 independent draws of $y$ according to Model~\ref{eqn:model}. Then, we split these 25 simulated datasets into 25 training and testing sets, where the training set consists of a random 80\% of the full set. Using these 25 training datasets, we estimate the posterior distribution of our parameters, from Formula~\ref{eqn:posterior_dist}, yielding 25 posterior distributions $\{(\alpha^{(j)}, \beta^{(j)}, \eta^{(j)})\}_{j=1}^{25}$. Using these 25 testing sets, we evaluate the performance of Model~\ref{eqn:model} by computing several metrics with these posterior distributions, detailed in Section~\ref{sec:simResults}.

%%%%%%%%%%%%%%%%%%%%%%%%%%%%%%%%%%%%%%%%%%%%%%%%%%%%%%%%%%%%%%%%%%%%%%%%%%%%%%%%%%%%%%%%%%%%%%%%%%%%%%%%%%%%%%

\textbf{Simulation results.}
In all three simulation studies, we reliably recover the data generating parameters: averaged across all parameters, the estimated frequentist coverage of the marginal 95\% posterior credible intervals exceeds 92\% in each study. 
Importantly, the coverage of the 95\% posterior credible intervals for the discontinuity parameters $\beta_{2k}$ and $\beta_{3k}$ exceeds 91\% in each study.
That is, for each simulated dataset, the 95\% credible intervals for the $\beta_{2k}$'s and $\beta_{3k}$'s usually contain the true data generating parameters.
Furthermore, our model demonstrates good predictive capabilities (see Appendix~\ref{app:simulation_details} for details).

\begin{figure}[hbt!]
\begin{subfigure}{0.32\textwidth}
\centering
\includegraphics[width = \textwidth]{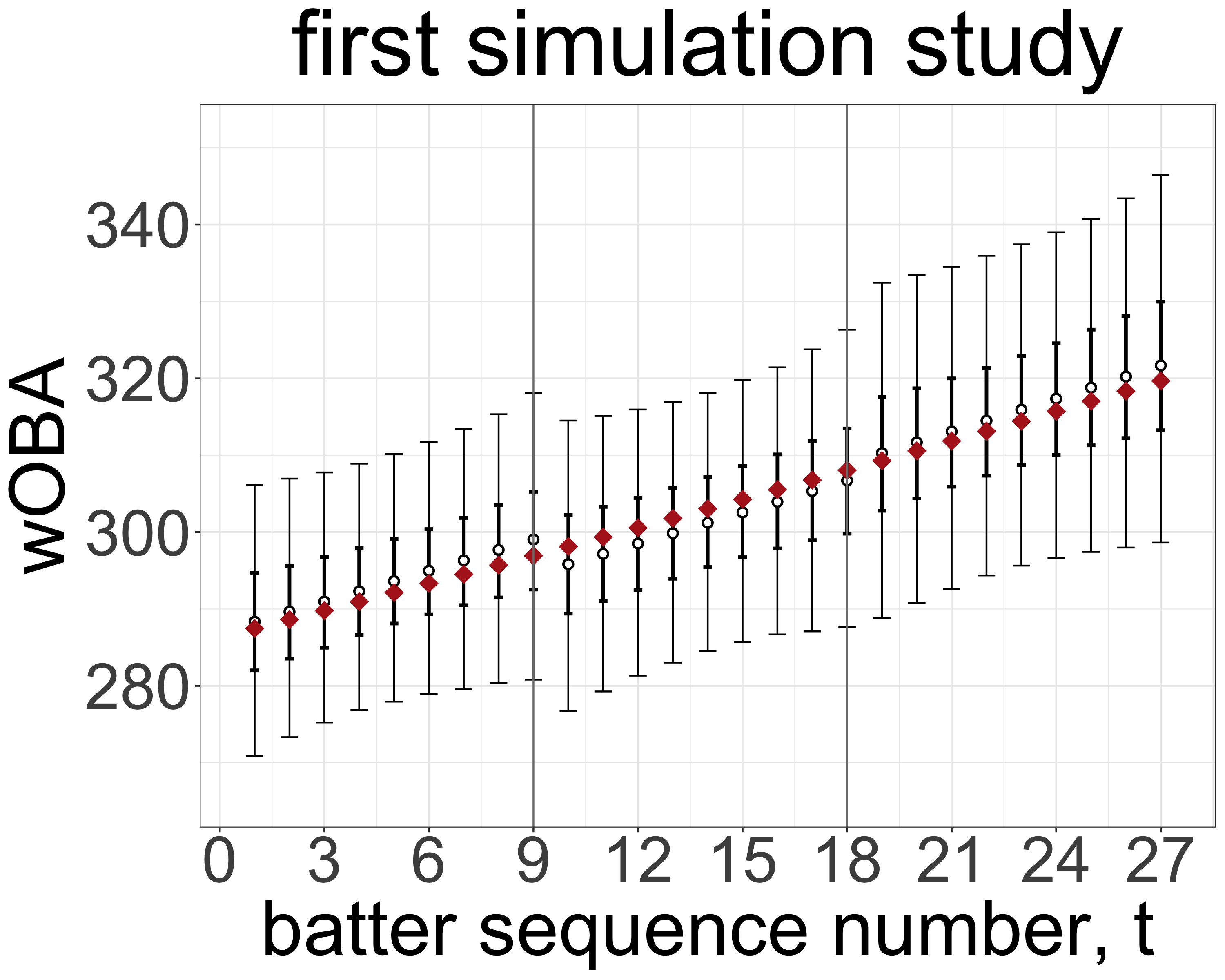}
% \caption{}
\label{fig:XwobaSim1}
\end{subfigure}
\begin{subfigure}{0.32\textwidth}
\centering
\includegraphics[width = \textwidth]{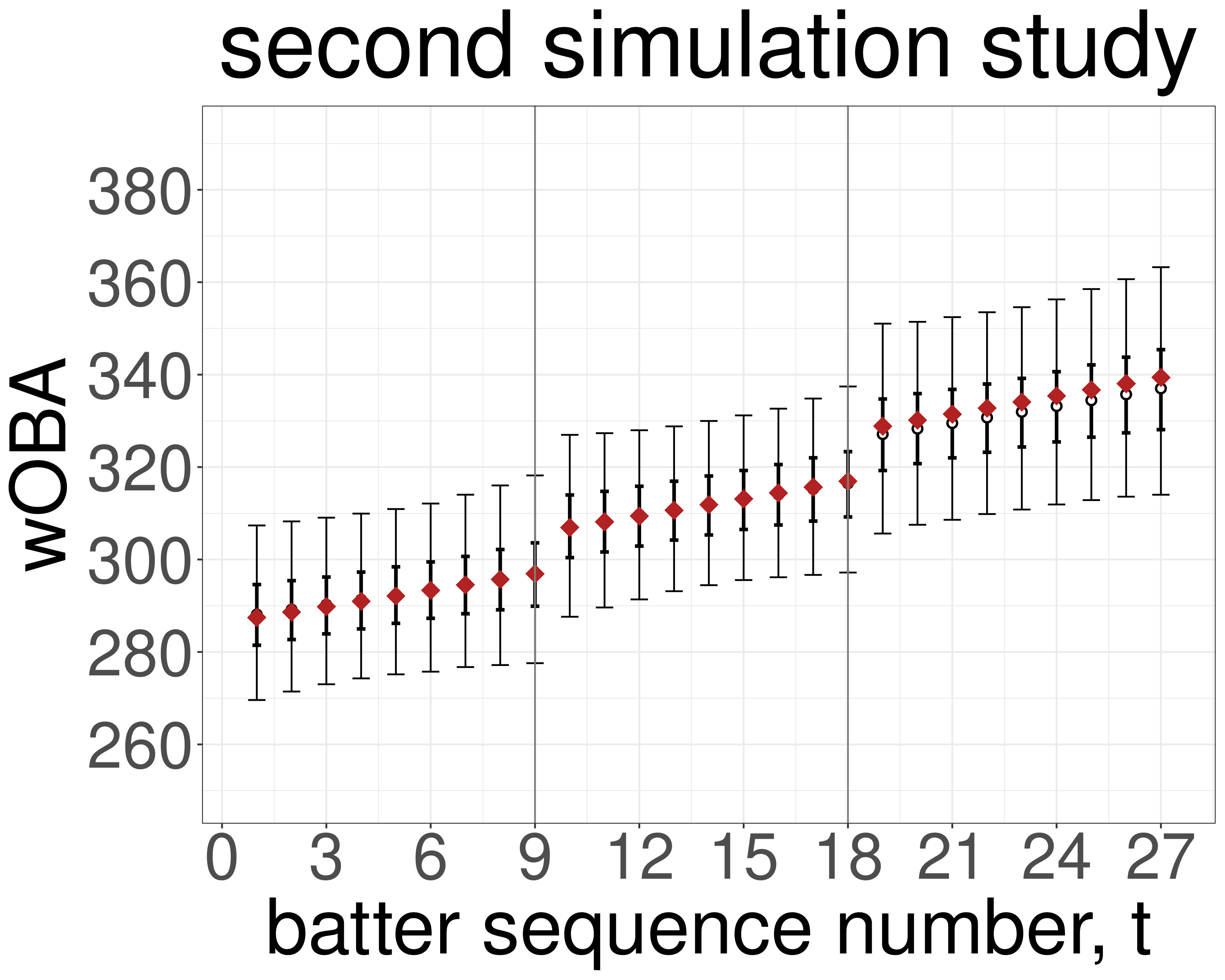}
% \caption{}
\label{fig:XwobaSim2}
\end{subfigure}
\begin{subfigure}{0.32\textwidth}
\centering
\includegraphics[width = \textwidth]{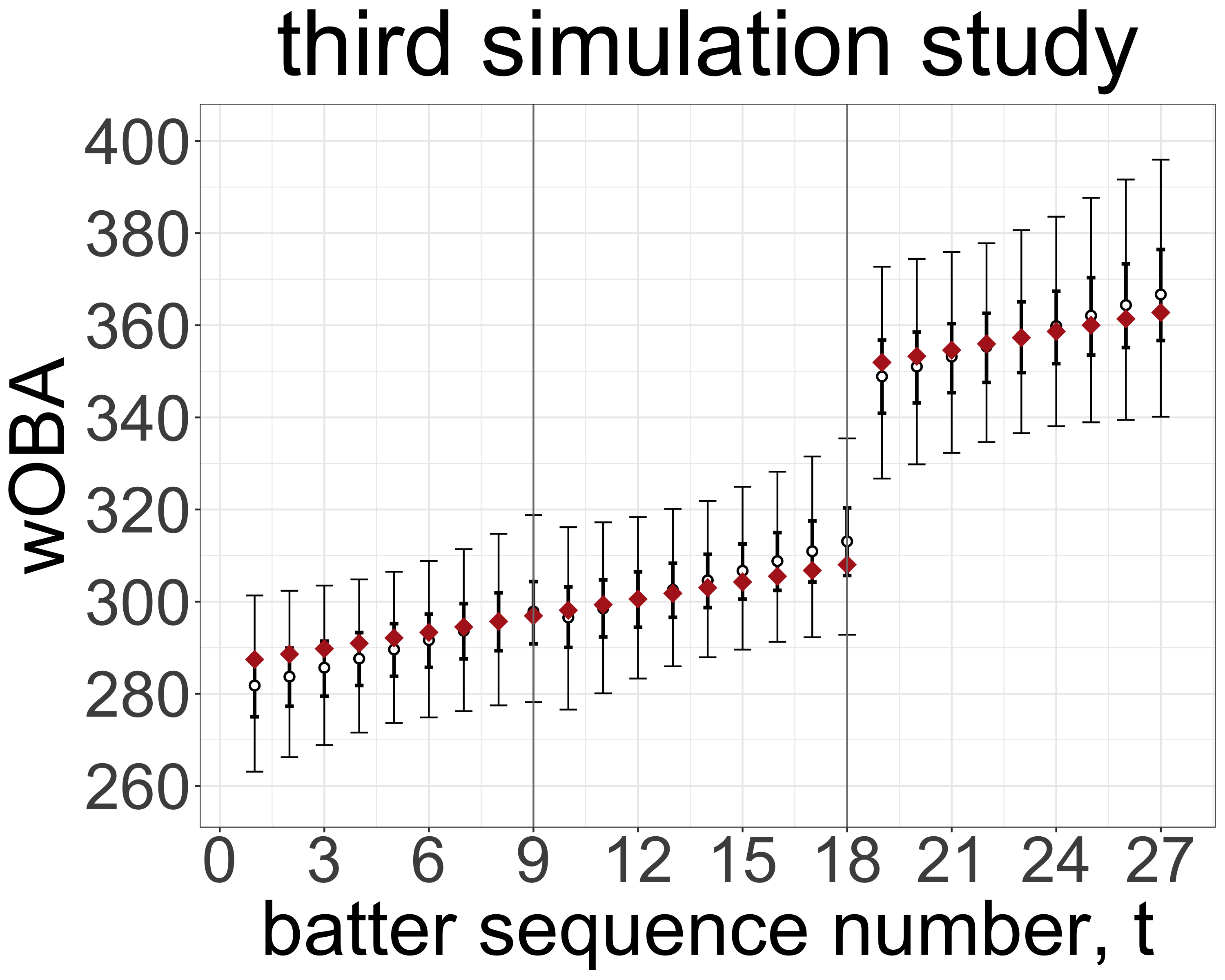}
% \caption{}
\label{fig:XwobaSim3}
\end{subfigure}
    \caption{Trend in $\xwoba$ over the course of a game from our first, second, and third simulation studies.
    The red dots indicate the true underlying expected $\woba$ values, the white dots indicate the posterior means of the $\xwoba$ values, the thick black error bars denote the 50\% posterior credible intervals, and the thin black error bars denote the 95\% posterior credible intervals.}
    \label{fig:plotXwobaSims}%
\end{figure}

%%%%%%%%%%%%%%%%%%%%%%%%%%%%%%%%%%%%%%%%%%%%%%%%%%%%%%%%%%%%%%%%%%%%%%%%%%%%%%%%%%%%%%%%%%%%%%%%%%%%%%%%%%%%%%
\textbf{Simulation visualization.}
In each simulation, we visualize the trajectory of posterior expected $\woba$ over the course of the game for an average batter on the road facing an average pitcher with the same handedness.
That is, we plot the sequence $\{\xwoba(t,\tilde{\bx})\}_{t = 1}^{27}$ where 
\begin{equation}
\label{eqn:x_tilde2}
% \tilde{\bx}^\top = (\text{logit}(0.315), \text{logit}(0.315), 1, 0).
\tilde{\bx}^\top = (\overline{x^{(b)}}, \overline{x^{(p)}}, 1, 0).
\end{equation}
Figure~\ref{fig:plotXwobaSims} shows the sequence of posterior means, 50\%, and 95\% credible intervals of $\xwoba(t,\tilde{\bx})$ based on a single simulated dataset from each simulation setting.
We overlay the true values of $\xwoba(t,\tilde{\bx}),$ computed from the data generating parameters, to each plot. 
We see that in each of the three simulation studies, we recover the true underlying expected $\woba$  trajectory. 
% Although there is some bias in the second simulation, the magnitude of the bias is small, and the true expected $\woba$  values lie in the 50\% credible intervals.

%%%%%%%%%%%%%%%%%
%%%%%%%%%%%%%%%%%

%%%%%%%%%%%%%%%%%%%%%%%%%%%%%%%%%%%%%%%%%%%%%%%%%%%%%%%%%%%%%%%
%%%%%%%%%%%%%%%%%%%%%%%%%%%%%%%%%%%%%%%%%%%%%%%%%%%%%%%%%%%%%%%
\section{Simulation details}
\label{app:simulation_details}

%%%%%%%%%%%%%%%%
\subsection{Data generating parameters}

The exact data generating parameter values of $\beta_{2k}$ and $\beta_{3k}$ for our three simulation studies are shown in Table~\ref{table:sim_betas}.

\begin{table}[H]
\centering
\caption{The data generating parameter values of $\beta_{2k}$ and $\beta_{3k}$ in each of our three simulations.}
\label{table:sim_betas}
\begin{tabular}{ l|llllll } \hline
 & $k=$ BB & $k=$ HBP & $k=$ 1B & $k=$ 2B & $k=$ 3B & $k=$ HR \\ \hline
$\beta_{2k}$ for sim 1 & 0 & 0 & 0 & 0 & 0 & 0 \\
$\beta_{3k}$ for sim 1 & 0 & 0 & 0 & 0 & 0 & 0 \\
$\beta_{2k}$ for sim 2 & 2/65 & 0 & 4/65 & 2/65 & 0 & 2/65 \\
$\beta_{3k}$ for sim 2 & 1/15 & 0 & 2/15 & 1/15 & 0 & 1/15 \\
$\beta_{2k}$ for sim 3 & 0 & 0 & 0 & 0 & 0 & 0 \\
$\beta_{3k}$ for sim 3 & 1/10 & 1/10 & 3/10 & 1/10 & 1/10 & 3/20 \\
\hline
\end{tabular}

\end{table}

Furthermore, in each of our simulation studies, we assume that pitchers fatigue linearly over the course of a game. The particular true parameter values of $\alpha_{0k}$ and $\alpha_{1k}$ used in each of our simulation studies are shown in Table~\ref{table:sim_alphas}.
\begin{table}[H]
\centering
\caption{The data generating parameter values of $\alpha_{0k}$ and $\alpha_{1k}$ in each of our three simulations.}
\label{table:sim_alphas}
\begin{tabular}{ l|llllll } \hline
 & $k=$ BB & $k=$ HBP & $k=$ 1B & $k=$ 2B & $k=$ 3B & $k=$ HR \\ \hline
$\alpha_{0k}$ & -0.601 & -1.804 & -0.475 & -0.943 & -1.510, & -0.565 \\
$\alpha_{1k}$ & 0.00271 & 0.0122 & 0.00354 & 0.00635 & 0.0223 & 0.00926 \\
\hline
\end{tabular}

\end{table}

Finally, in each of our simulation studies, we set the value of $\eta$ to mimic fitted values from observed data. The particular true parameter values of $\eta$ used in each of our simulation studies are shown in Table~\ref{table:sim_etas}.
\begin{table}[H]
\centering
\caption{The data generating parameter values of $\eta$ in each of our three simulations.}
\label{table:sim_etas}
\begin{tabular}{ l|llllll } \hline
 & $k=$ BB & $k=$ HBP & $k=$ 1B & $k=$ 2B & $k=$ 3B & $k=$ HR \\ \hline
$\eta_{\text{bat\_quality}}$ & 0.865 & 1.408 & 0.371 & 0.856 & 1.399, & 1.525 \\
$\eta_{\text{pit\_quality}}$ & 1.128 & 1.987 & 1.050 & 1.472 & 3.286 & 1.850 \\
$\eta_{\text{hand}}$ & -0.201 & 0.166 & -0.0164 & -0.0420 & -0.462 & -0.0958 \\
$\eta_{\text{home}}$ & 0.0792 & -0.0776 & 0.0245 & -0.00103 & 0.107 & 0.0230 \\
\hline
\end{tabular}
\end{table}

%%%%%%%%%%%%%%%%
\subsection{Predictive performance on simulated data}

Our model demonstrates good predictive capabilities. 
%In Section~\ref{sec:model_sim}, we discussed that our model exhibits good parameter coverage. Here, we we report out-of-sample cross entropy loss.
To get a general sense of our model's performance, we use out-of-sample cross entropy loss, given by
\begin{equation}
\label{eqn:cross_entropy_loss}
 - \frac{1}{n} \sum_{i=1}^{n} \sum_{k=1}^{7} 1\{y_i=k\} \cdot \log \big( \P(y_i = k) \big) .
\end{equation}
For each of our three simulations, the average cross entropy loss over each of our 25 datasets is 1.05, 1.06, and 1.07, respectively.
% Note, simply predicting that each plate appearance occurs with probability 1/7 yields an average cross-entropy loss of $\log(7) = 1.945.$
Using the empirical outcome probabilities yields an average out-of-sample cross-entropy loss of 1.06, 1.08, and 1.08, respectively, for each of our three simulations.
It is reassuring that our model (barely) outperforms the observed base rates. %and the naive uniform probabilities.
 
%We compare our model's cross entropy loss to that of other prediction strategies to better understand its performance. 
%First, consider using a uniform distribution to predict the probability of each plate appearance outcome, assigning each outcome a probability of $1/7$. The cross entropy loss of the uniform distribution on 7 outcomes is $\log 7 = 1.945$. Our model outperforms the uniform distribution. 
%Second, consider using the base rates of each plate appearance outcome to predict the probability of each outcome (e.g., the proportion of plate appearances in which a single occurs to predict the probability that a single occurs). The average cross entropy loss over each of our 25 datasets, using the base rates, is 1.27. So, our model outperforms the base rates.

%%%%%%%%%%%%%%%%%%%%%%%%%%%%%%%%%%%%%%%%%%%%%%%%%%%%%%%%%%%%%%%
%%%%%%%%%%%%%%%%%%%%%%%%%%%%%%%%%%%%%%%%%%%%%%%%%%%%%%%%%%%%%%%
\section{Observed model fit details}\label{app:effect_sizes}

%%%%%%%%%%%%%%%%%%%%%%%%%%%%%%%%%%%%%%%%%%%%%%%%%%%%%%%%%%%%%%%%%%%%%%%%%%%%%%
\subsection{The impact of pitcher decline on the outcome of a plate appearance}\label{sec:prob_scale_TTO}  

In this Section, we quantify the effect size of pitcher decline over the course of a game, again using the 2017 season as our primary example. 

In particular, we examine how the probability of each outcome of a plate appearance changes over the course of a game. Specifically, we use the posterior distribution of $\P(y = k \vert t,\bx)$, defined in Formula~\eqref{eqn:prob_ktx}, to characterize the amount by which pitchers decline within a game. In particular, we compute the posterior distribution of the change in the probability of outcome $k \neq 1$ from $\ttoone$ to $\ttotwo$, over average,
\begin{equation}
\label{eqn:D12kx}
\mathscr{D}_{12}(k,\bx) = \frac{1}{9} \sum_{t=10}^{18} \P(y = k|t,\bx) - \frac{1}{9} \sum_{t=1}^{9} \P(y = k|t,\bx), 
\end{equation}
and the similarly defined $\mathscr{D}_{23}(k,\bx)$, which captures the change in the probability of outcome $k \neq 1$ from $\ttotwo$ to $\ttothree$, over average.

\begin{figure}[H]
\centering
\includegraphics[width=15cm]{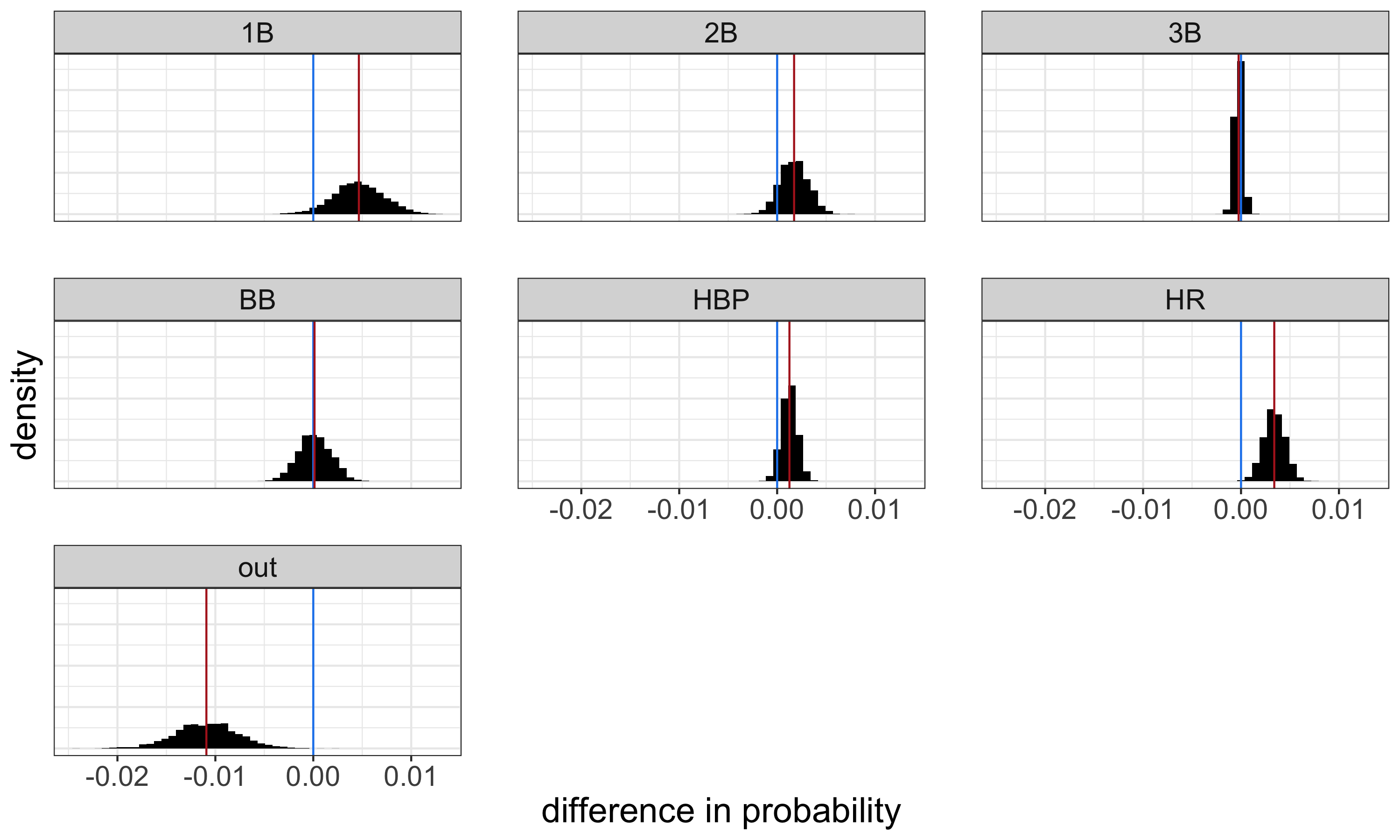}
\caption{The difference in probability of each plate appearance outcome between $\ttotwo$ and $\ttoone$ on average (assuming a batter of average quality on the road faces a pitcher of average quality with a handedness match during each plate appearance). Equivalently, the posterior distribution of $\mathscr{D}_{12}(k,\tilde{\bx})$. The red line denotes the mean, and the blue line denotes 0.}
\label{fig:tto_diff_12_x1}
\end{figure}

In Figure~\ref{fig:tto_diff_12_x1} we plot the posterior distribution of $\mathscr{D}_{12}(k,\tilde{\bx})$, using plate-appearance-state vector $\tilde{\bx}$ from Formula~\eqref{eqn:x_tilde}.
% The probability of a single increases by about 0.0075 from $\ttoone$ to $\ttotwo$, and the probability of the other categories change negligibly. 
From $\ttoone$ to $\ttotwo$, the probability of a single increases by about 0.005, the probability of a home run increases by about 0.003, and the probability of the other non-out categories change negligibly. 
With this, the probability of an out decreases by about 0.01.
So, there is a small decrease in pitcher performance on average from $\ttoone$ to $\ttotwo$.

\begin{figure}[H]
\centering
\includegraphics[width=15cm]{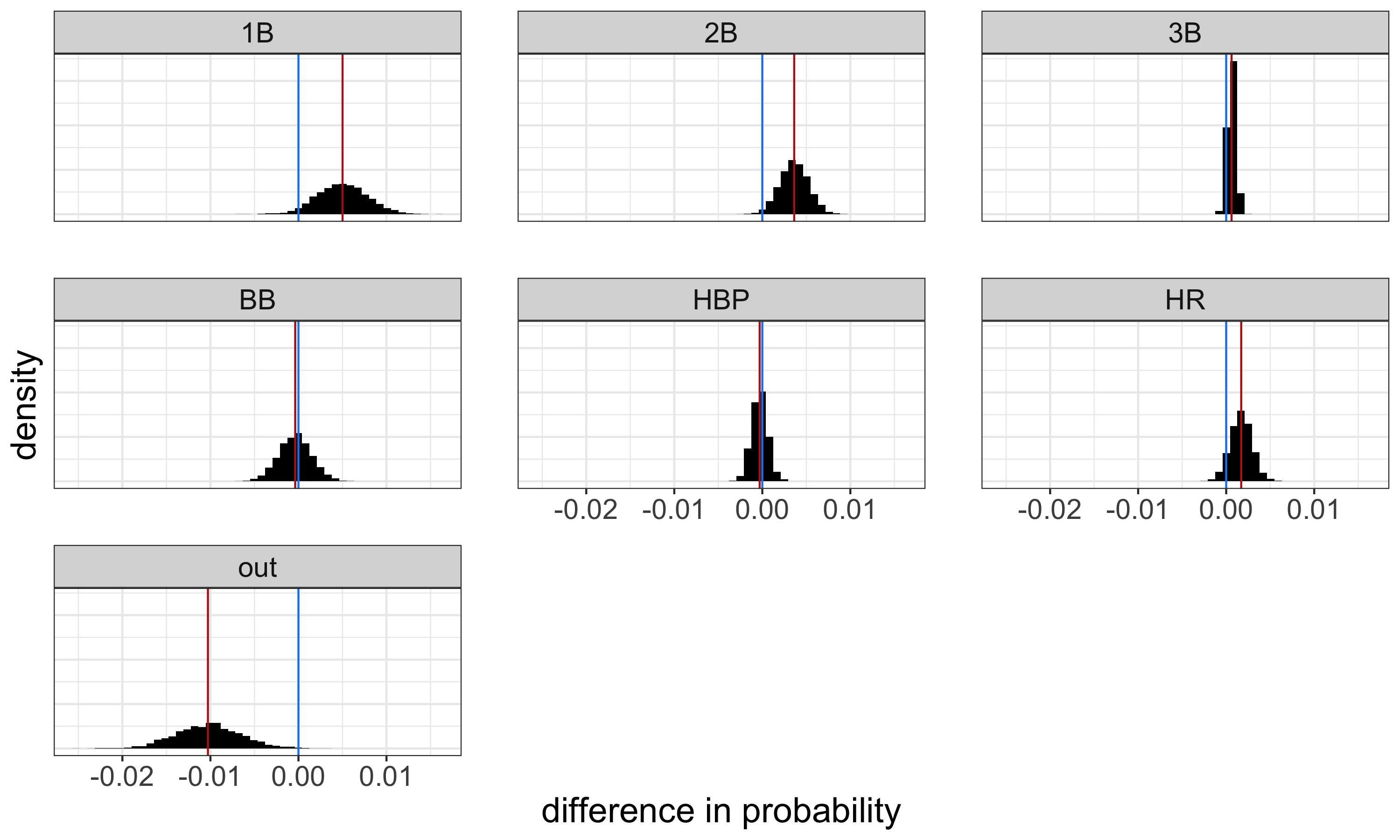}
\caption{The difference in probability of each plate appearance outcome between $\ttothree$ and $\ttotwo$ on average (assuming a batter of average quality on the road faces a pitcher of average quality with a handedness match during each plate appearance). Equivalently, the posterior distribution of $\mathscr{D}_{23}(k,\tilde{\bx})$.  The red line denotes the mean, and the blue line denotes 0.} 
\label{fig:tto_diff_23_x1}
\end{figure}

Similarly, in Figure~\ref{fig:tto_diff_23_x1} we plot the posterior distribution of $\mathscr{D}_{23}(k,\tilde{\bx})$.
From $\ttotwo$ to $\ttothree$, the probability of a single increases by about 0.005, the probability of a double increases by about 0.004, 
% the probability of a home run increases by about 0.005, 
and the probability of the other non-out categories change negligibly. 
With this, the probability of an out decreases by about 0.01.
So, there is a small decrease in pitcher performance on average from $\ttotwo$ to $\ttothree$.

Additionally, we examine how the expected $\woba$ of each outcome of a plate appearance changes over the course of a game.
In particular, we compute the posterior distribution of the change in the expected $\woba$ of outcome $k \neq 1$ from $\ttoone$ to $\ttotwo$, over average, 
\begin{equation}
\label{eqn:D12kx_xw}
\mathscr{D}'_{12}(k,\bx) = \frac{1}{9} \sum_{t=10}^{18} 1000 \cdot w_k\cdot\P(y = k|t,\bx) - \frac{1}{9} \sum_{t=1}^{9} 1000 \cdot w_k\cdot\P(y = k|t,\bx), 
\end{equation}
where $w_k$ is the $\woba$ weight for outcome $k$ as discussed in Section~\ref{sec:wOBA}.
Similarly, we define $\mathscr{D}'_{23}(k,\bx)$, which captures the change in the expected $\woba$ of outcome $k \neq 1$ from $\ttotwo$ to $\ttothree$, over average.

\begin{figure}[H]
\centering
\includegraphics[width=15cm]{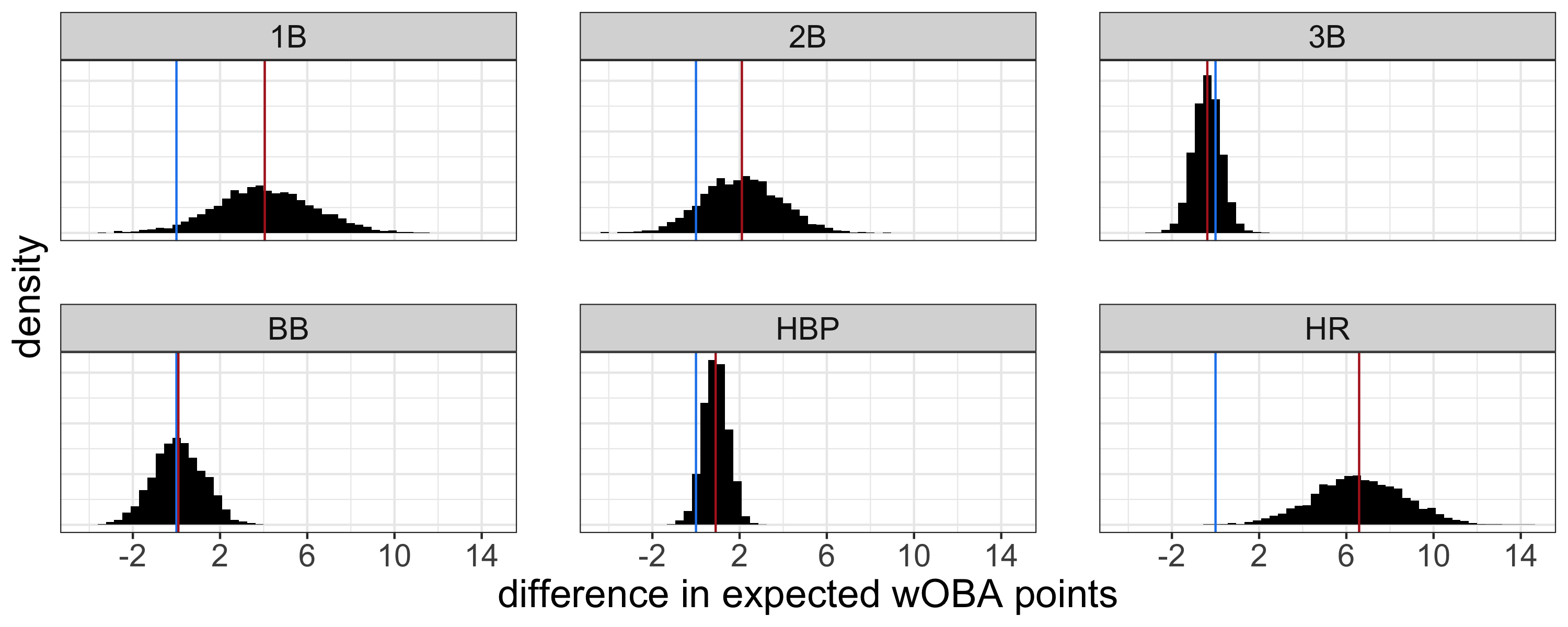}
\caption{The difference in $\xwoba$ of each plate appearance outcome between $\ttotwo$ and $\ttoone$ on average (assuming a batter of average quality on the road faces a pitcher of average quality with a handedness match during each plate appearance). Equivalently, the posterior distribution of $\mathscr{D}'_{23}(k,\tilde{\bx})$.  The red line denotes the mean, and the blue line denotes 0.} 
\label{fig:tto_xw_diff_12_x1}
\end{figure}

In Figure~\ref{fig:tto_xw_diff_12_x1} we plot the posterior distribution of $\mathscr{D}'_{12}(k,\tilde{\bx})$, using plate-appearance-state vector $\tilde{\bx}$ from Formula~\eqref{eqn:x_tilde}.
From $\ttoone$ to $\ttotwo$, the expected $\woba$ points of a home run increases by about six, the expected $\woba$ points of a single increases by about four,
 % the expected $\woba$ points of a double increases by about two,
 and the other non-out categories change negligibly. 
 Note that the expected $\woba$ of an out doesn't change because an out is worth zero $\woba$.
% So, there is a small decrease in pitcher performance on average from $\ttoone$ to $\ttotwo$.

\begin{figure}[H]
\centering
\includegraphics[width=15cm]{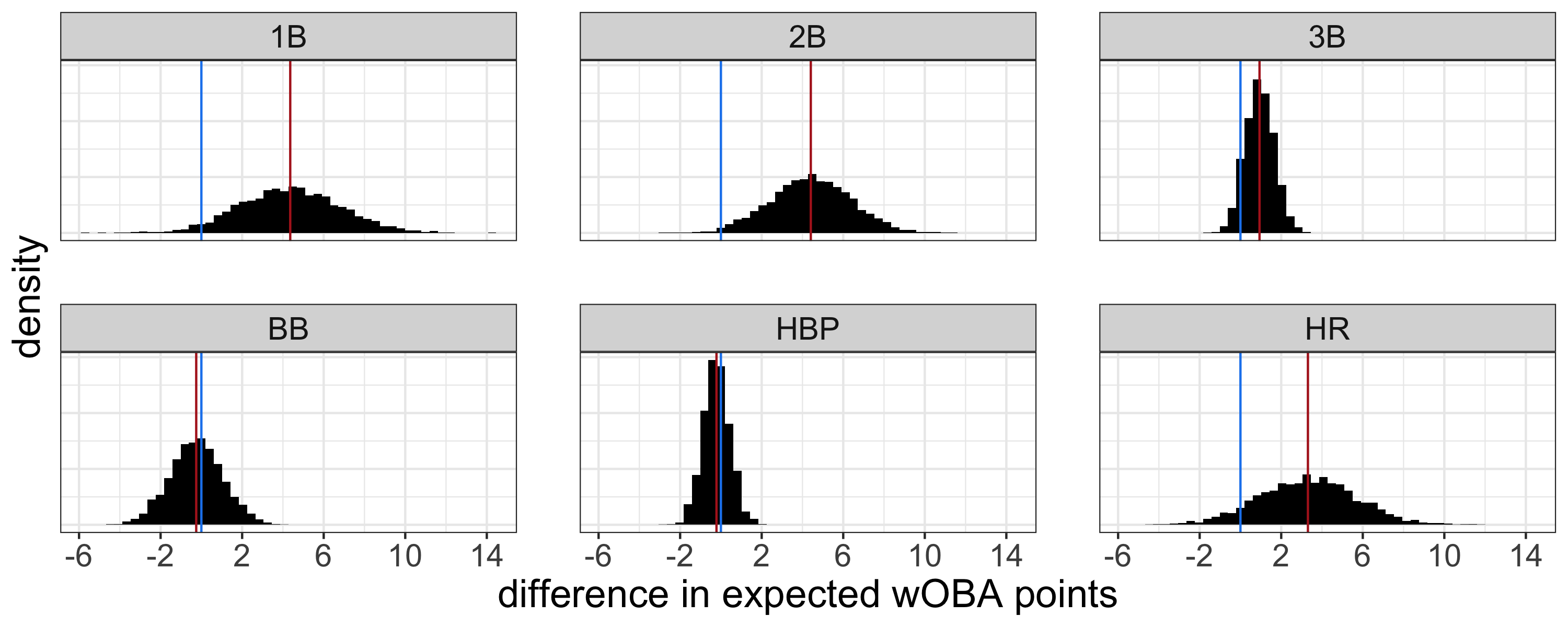}
\caption{The difference in $\xwoba$ of each plate appearance outcome between $\ttothree$ and $\ttotwo$ on average (assuming a batter of average quality on the road faces a pitcher of average quality with a handedness match during each plate appearance). Equivalently, the posterior distribution of $\mathscr{D}'_{23}(k,\tilde{\bx})$.  The red line denotes the mean, and the blue line denotes 0.} 
\label{fig:tto_xw_diff_23_x1}
\end{figure}

Similarly, in Figure~\ref{fig:tto_xw_diff_23_x1} we plot the posterior distribution of $\mathscr{D}'_{23}(k,\tilde{\bx})$.
From $\ttotwo$ to $\ttothree$, the expected $\woba$ of a double and single increases by about five, the $\xwoba$ of a home run increases by about three, and the other categories change negligibly. 
% So, there is a small decrease in pitcher performance on average from $\ttotwo$ to $\ttothree$.

%%%%%%%%%%%%%%%%%%%%%%%%%%%%%%%%%%%%%%%%%

\begin{figure}[h]

\begin{subfigure}{0.49\textwidth}
\centering
\includegraphics[width = \textwidth]{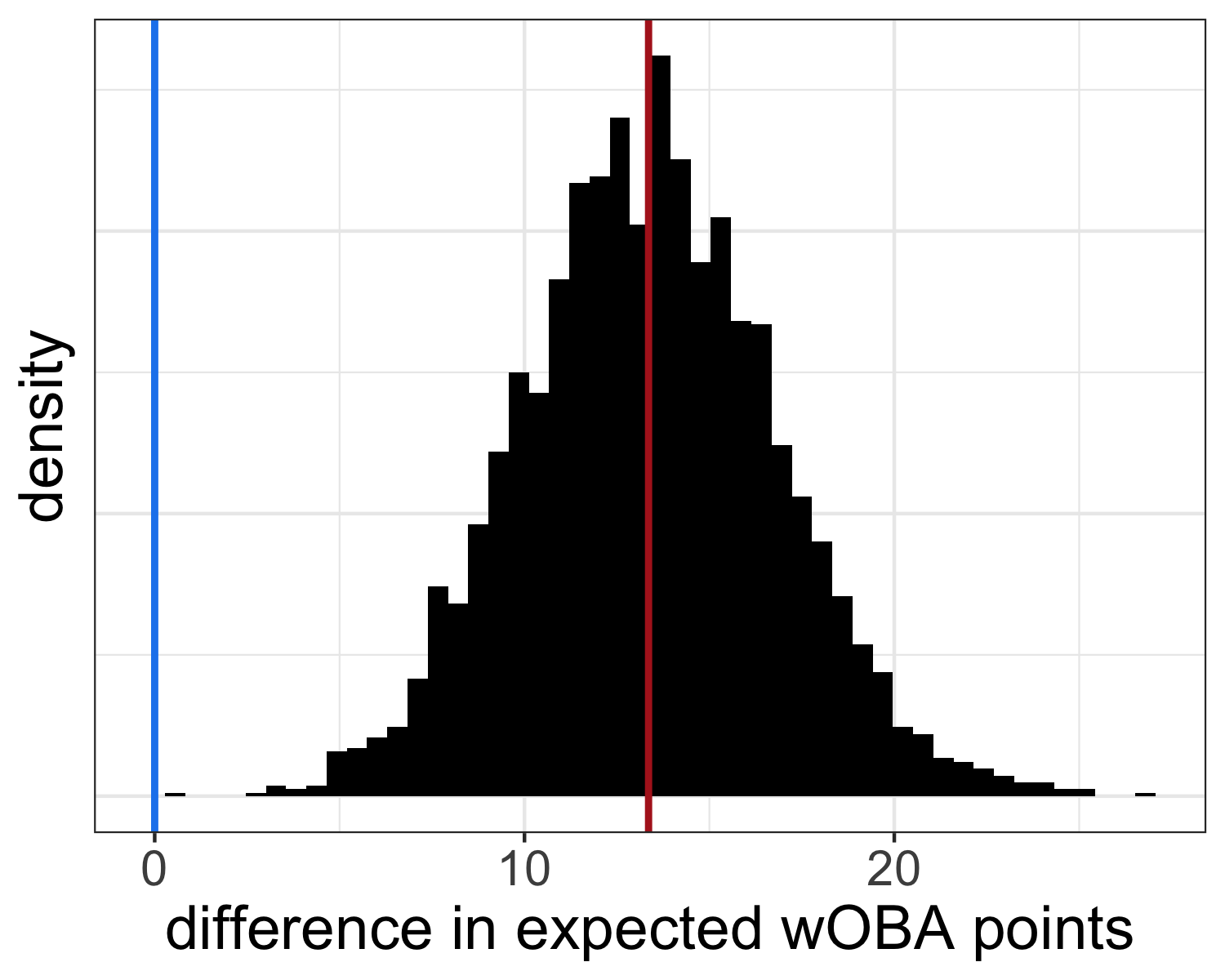}
\caption{}
\end{subfigure}
\begin{subfigure}{0.49\textwidth}
\centering
\includegraphics[width = \textwidth]{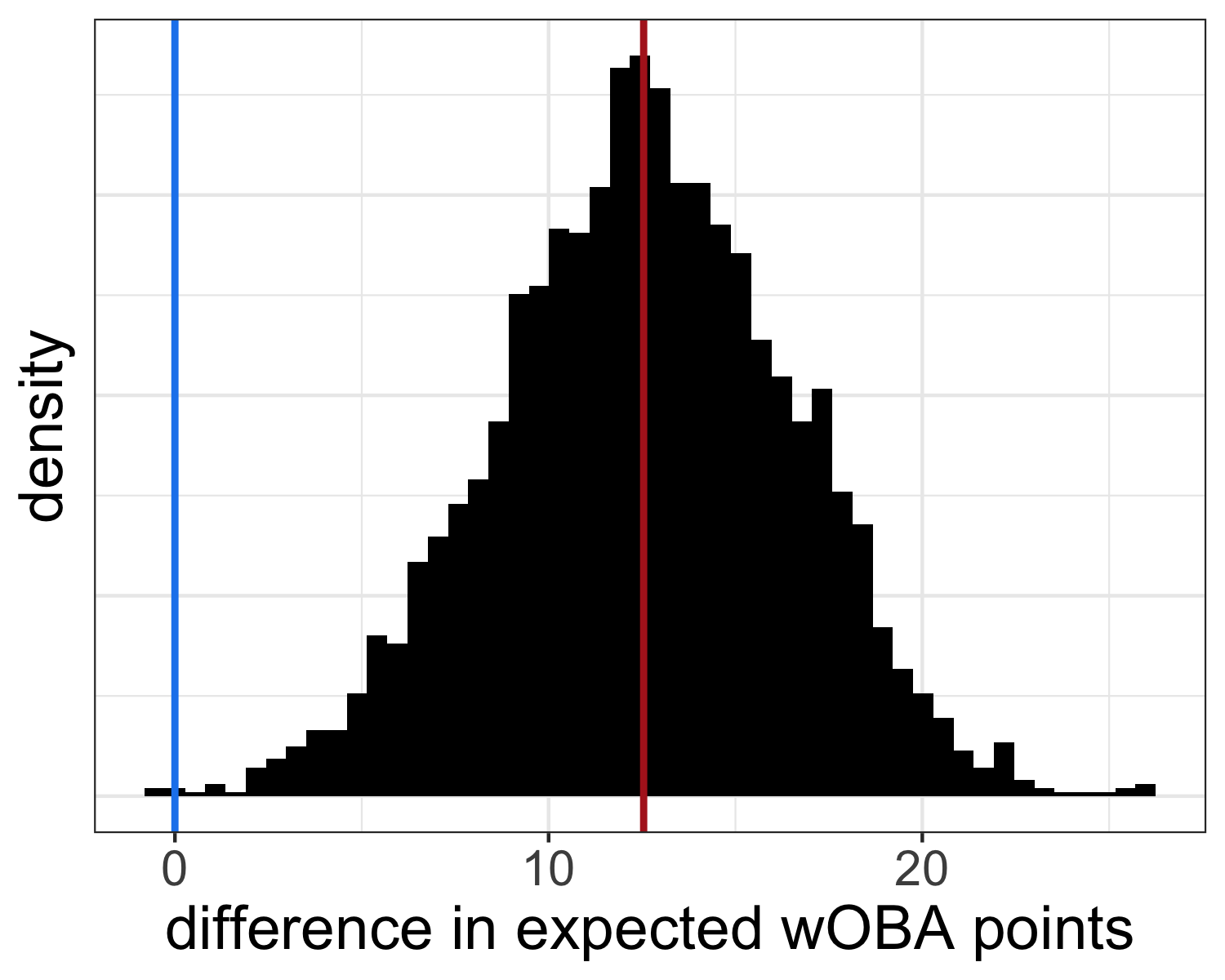}
\caption{}
\end{subfigure}
    \caption{The posterior distribution of the mean batter improvement, or mean pitcher decline in $\xwoba$, from $\ttoone$ to $\ttotwo$ (left) and from $\ttotwo$ to $\ttothree$ (right). 
    Equivalently, the posterior distributions of $\mathscr{D}_{12}(\tilde{\bx})$ (left) and $\mathscr{D}_{23}(\tilde{\bx})$ (right) (see Formula~\eqref{eqn:D12x}). The red line denotes the mean, and the blue line denotes 0. We see that batters improve relative to the pitcher by about 13 $\woba$ points on average from one $\tto$ to the next.}
    \label{fig:tto_diff_xw}
\end{figure}

Furthermore, we aggregate the increase in the probability of each non-out plate appearance outcome $k$ from one TTO to the next via expected $\woba$, defined in Equation~\eqref{eqn:xwOBA_tx}. 
In particular, recall from Section~\ref{sec:tango_comp} that a pitcher declines by about 13 $\woba$ points from one TTO to the next, over average, which is consistent with the effect sizes from Figures \ref{fig:tto_diff_12_x1} and \ref{fig:tto_diff_23_x1}. 
Figure~\ref{fig:tto_diff_xw} illustrates this via a histogram of the posterior samples of $\mathscr{D}_{12}(\tilde{\bx})$ and $\mathscr{D}_{23}(\tilde{\bx})$.
We see that virtually all of these samples are positive, suggesting that average pitcher performance declines from one $\tto$ to the next, and that the means of these distributions are around 13 $\woba$ points, which are consistent with \citet{theBook}'s findings. 
Specifically, our model suggests that the expected $\woba$ points of an average plate appearance increases by $13.4$ (with a $95\%$ credible interval of [7.78, 19.0]) from the first $\tto$ to the second, and by $12.5$ (with a $95\%$ credible interval of [5.98, 18.7]) from the second $\tto$ to the third. 

%%%%%%%%%%%%%%%%%
\subsection{Predictive performance on observed data}\label{sec:model_performance}
To get a general sense of our model's performance on observed data, we run a five-fold cross validation to predict the probability of each plate appearance outcome for each plate appearance in 2017. 
The out-of-sample cross entropy loss, given by Formula~\eqref{eqn:cross_entropy_loss}, is 1.035. 
We compare our model's cross entropy loss to that of other prediction strategies to better understand its performance. 
% First, consider using a uniform distribution to predict the probability of each plate appearance outcome, assigning each outcome a probability of $1/7$. 
% The cross entropy loss of the uniform distribution on 7 outcomes is $\log 7 = 1.945$. 
% Our model outperforms the uniform distribution, which makes sense because the distribution of the outcome of a plate appearance is far from uniform (e.g., there are many more outs than home runs).
% Second, consider a five-fold cross validation using the base rates of each plate appearance outcome. 
Consider a five-fold cross validation using the base rates of each plate appearance outcome. 
So, for each fold, find the proportion of plate appearances in which each outcome occurs, and compute the cross entropy loss using these base rates on the remaining out-of-sample plate appearances. 
For reference, in 2017, an out occurs in 67.6\% of plate appearances, an uBB 7.8\%, a HBP 0.9\%, a 1B 14.9\%, a 2B 4.8\%, a 3B 0.45\%, and a HR in 3.5\% of plate appearances. 
The out-of-sample cross entropy loss of the base rates of each outcome is 1.042. 
So, our model very slightly outperforms the base rates.
Finally, note that our model using raw batter and pitcher quality covariates, rather than logit-transformed batter and pitcher quality covariates, has a cross-validated out-of-sample cross entropy loss of 1.040.
That the logit-transformed player quality covariates have better out-of-sample predictive performance helps justify using the logit transform.

\subsection{The trend is persistent across years}\label{sec:trend_across_years}  

\begin{figure}[hbt!]
\centering
\includegraphics[width=17cm]{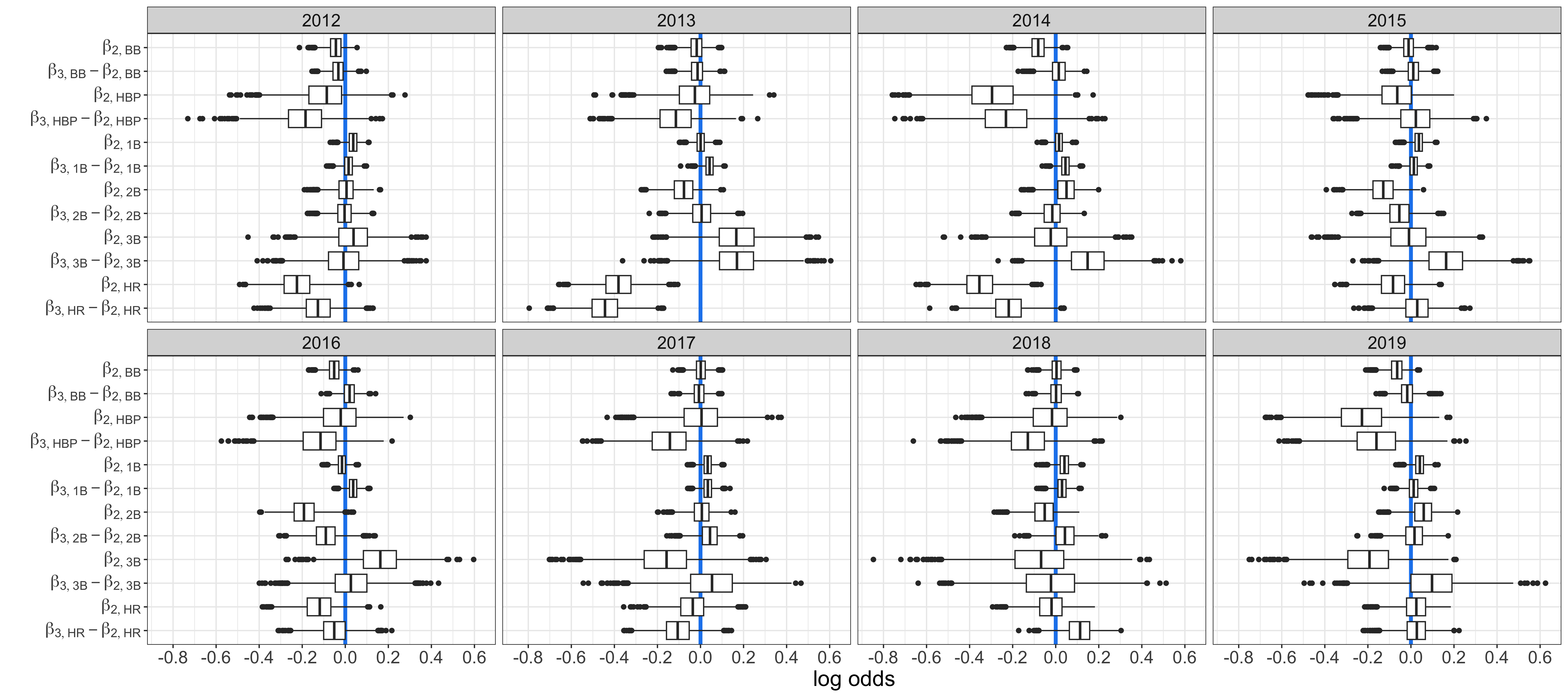}
\caption{
Posterior boxplots of the $\ttop$ discontinuity parameters from Model~\eqref{eqn:model}, fit separately on data from each year from 2012 to 2019. The blue line denotes 0. We see that each posterior distribution covers both positive and negative values.
} 
\label{fig:beta_boxplot_ALLYRS}
\end{figure}

In Figure~\ref{fig:beta_boxplot_ALLYRS} we show boxplots of the posterior distributions of the discontinuity parameters $\beta_{2k}$ and $\beta_{3k} - \beta_{2k}$ from our model (Equation~\eqref{eqn:model}) fit separately on data from each season from 2012 to 2019.
For some outcomes (e.g. walks), the posterior distributions are tightly concentrated around 0, and for other outcomes (e.g., triples and hit-by-pitches, which are rare events), the posterior distributions are quite wide, which is compatible with a large effect in either direction.
Overall, the posterior distributions of the discontinuity parameters cover both positive and negative values, and most of them are centered around 0.
In particular, we don't see what we would expect to see if there were strong evidence for a $\ttop$ (i.e., we don't see the posterior distributions tighly concentrated around a positive number).
Ultimately, we do not find the posterior distributions in Figure~\ref{fig:beta_boxplot_ALLYRS} to be consistent with large, systematic time through the order penalties.
% Thus we don't find overwhelming evidence of strong discontinuity in the probability of each plate appearance outcome.

%%%%%%%%%%%%%%%%%%%%%
%%%%%%%%%%%%%%%%%%%%%

\begin{figure}[hbt!]
\centering
\includegraphics[width=17cm]{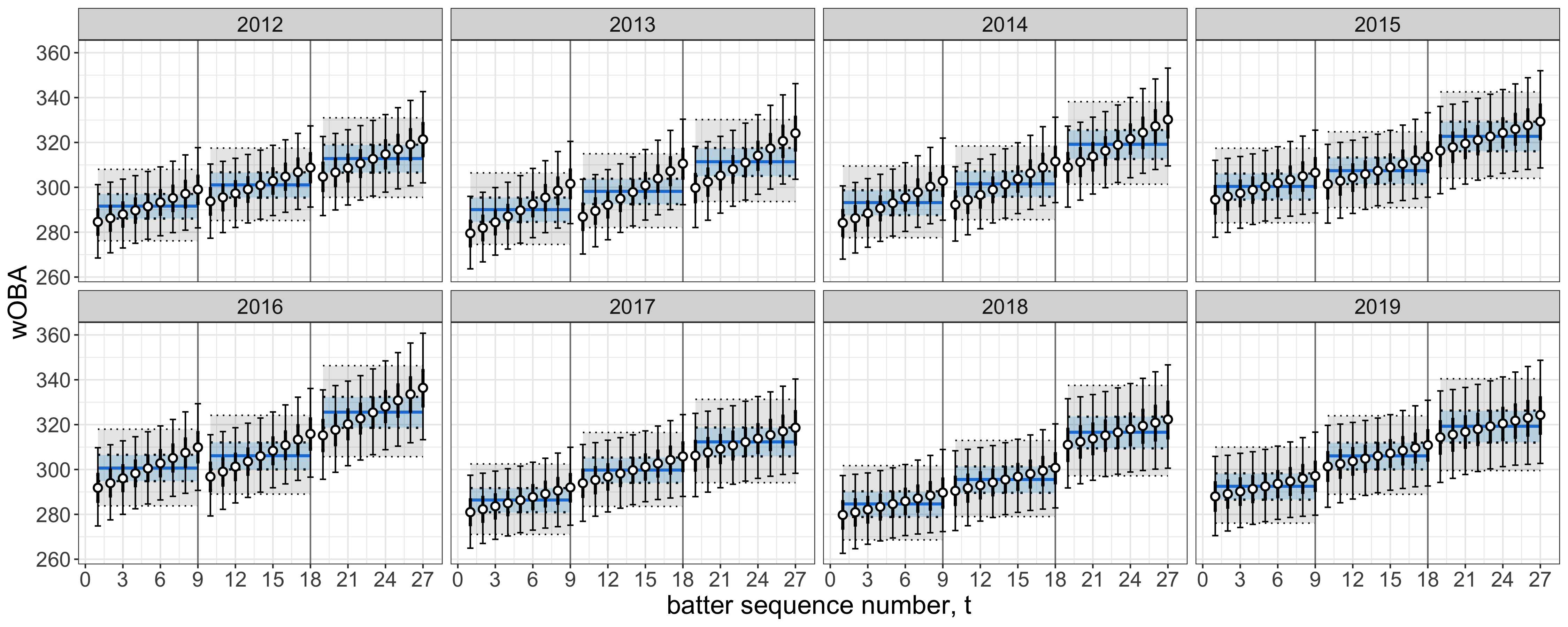}
\caption{
Trend in expected $\woba$ over the course of a game for an average batter facing an average pitcher of the same handedness on the road, according to the model from Equation~\eqref{eqn:model} fit on separately on data from each year from 2012 to 2019.
The white dots indicate the posterior means of the expected $\woba$ values, the thick black error bars denote the 50\% credible intervals, and the thin black error bars denote the 95\% credible intervals.} 
\label{fig:xwoba_trend_across_years}
\end{figure}

In Figure~\ref{fig:xwoba_trend_across_years} we plot the posterior distribution of $\xwoba$ over the course of a game according to our model fit separately on data from each year from 2012 to 2019. 
% We see that $\xwoba$ increases steadily over the course of a game.
% Although some years see slight discontinuity between successive times through the order
We see that expected $\woba$ increases steadily over the course of a game, without significant discontinuity (in particular, significant \textit{upward} discontinuity) between times through the order.
The 2018 season is the only season in which we see an upward discontinuity in the posterior means, which occurs between $\ttotwo$ and $\ttothree$.
This discontinuity, however, lies inside of the credible intervals and so is not significant.

%%%%%%%%%%%%%%%%%%%%%
%%%%%%%%%%%%%%%%%%%%%

%%%%%%%%%%%%%%%%%%%%%%%%%%%%%%%%%%%%%%%%%%%%%%%%%%%%%%%%%%%%%%%
%%%%%%%%%%%%%%%%%%%%%%%%%%%%%%%%%%%%%%%%%%%%%%%%%%%%%%%%%%%%%%%
\section{Alternative models}\label{app:alt_models}

%%%%%%%%%%%%%%%%%%%%%%%%%%%%%%%%%%%%%%%%%%%%%%%%%%%%%%%%%%%%%%%%%%%%%%%%%%%%%%
\subsection{A more flexible model: the indicator model}\label{sec:indicator_model}  

In Equation~\eqref{eqn:model} we model pitcher decline over the course of a game as the combination of discontinuous decline from each $\tto$ to the next and continuous linear pitcher decline across all the batters.
A more flexible model wouldn't enforce a particular functional form on within-game pitcher decline.
In particular, the most flexible model has a separate coefficient for each batter $t \in \{1,...,27\}$,
\begin{align}
\label{eqn:indicator_model}
\log\left(\frac{\P(y_{i} = k)}{\P(y_{i} = 1)}\right) &= \sum_{t=1}^{27} \alpha_{tk} \ind{t_{i} = t} + \bx_{i}^{\top}\eta_{k}.
\end{align} 
% The change in pitcher performance over the course of a game on the log odds scale is given by $\{\alpha_{tk}\}_{t=1}^{27}$.

\begin{figure}[H]
\centering
\includegraphics[width=0.7\textwidth]{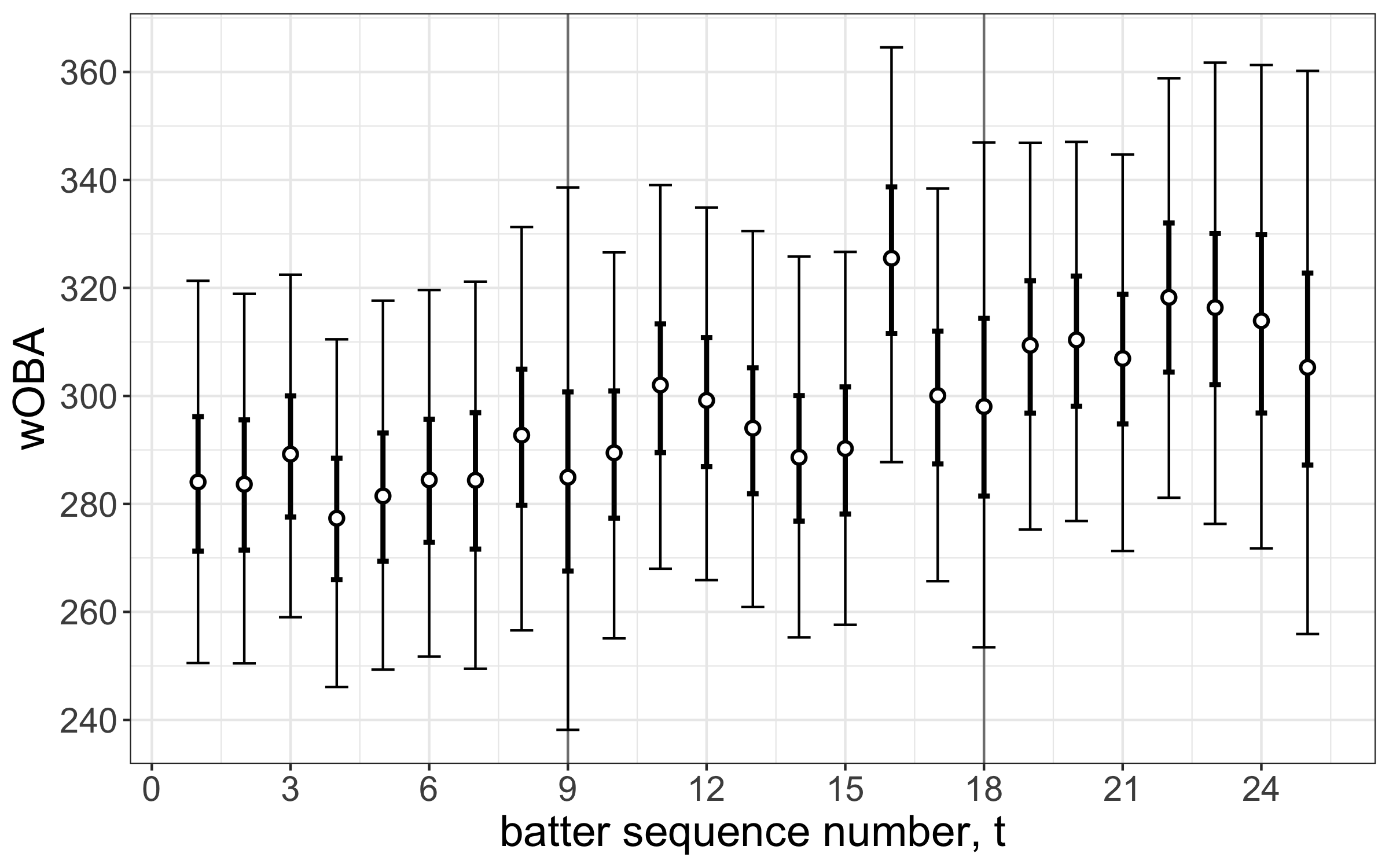}
\caption{
Trend in expected $\woba$ over the course of a game in 2017 for an average batter facing an average pitcher of the same handedness on the road, according to the indicator model from Equation~\eqref{eqn:indicator_model}.
The white dots indicate the posterior means of the expected $\woba$ values, the thick black error bars denote the 50\% credible intervals, and the thin black error bars denote the 95\% credible intervals.} 
\label{fig:plot_obs_results_2017_xwoba_check_indicator}
\end{figure}

With this more flexible model, the qualitative results of our study don't change.
For instance, as in Figure~\ref{fig:plot_xwoba_over_time_2017a}, in Figure~\ref{fig:plot_obs_results_2017_xwoba_check_indicator} we plot the posterior distribution of the trajectory of expected $\woba$ over the course of a game, according to the indicator model from Equation~\eqref{eqn:indicator_model} fit on data from 2017.
We do not see a significant discontinuity in pitcher performance from one $\tto$ to the next.
In other words, we don't find evidence of a strong batter discontinuity between times through the order.
% The qualitative results according to the indicator model are similar for other years.
This trend is persistent across each year from 2012 to 2019.

%expected $\woba$ increases steadily over the course of a game, without discontinuity due to batter learning in the second or third time through the order. 
% In other words, our model finds little evidence for an effect of batter learning on the expected $\woba$ of a plate appearance.
% This trend is persistent across each year from 2012 to 2019 and other choices of $\bx$.

%%%%%%%%%%%%%%%
%%%%%%%%%%%%%%%

%%%%%%%%%%%%%%%%%%%%%%%%%%%%%%%%%%%%%%%%%%%%%%%%%%%%%%%%%%%%%%%%%%%%%%%%%%%%%%
\subsection{A more elaborate model: pitcher-specific and batter-specific effects}\label{sec:varying_slopes_model}  

In our model from Equation~\eqref{eqn:model}, we make the simplifying assumption that the trajectory of within-game pitcher deterioration is the same across all pitchers and batters.
Nonetheless, it is likely that pitcher performance declines at different rates for different players.
To account for such heterogeneity, we extend our model by introducing player-specific rates of decline.
Specifically, we model
\begin{align}
\label{eqn:model_playerSpecificSlopes}
\log\left(\frac{\P(y_{i} = k)}{\P(y_{i} = 1)}\right) &= \alpha_{0kp(i)} + \alpha_{1kp(i)}t_{i} + \beta_{2kb(i)}\ind{t_{i} \in \ttotwo} + \beta_{3kb(i)}\ind{t_{i} \in \ttothree} + \bx_{i}^{\top}\eta_{k},
\end{align}
where $p(i)$ is the index of the pitcher and $b(i)$ is the index of the batter in at-bat $i$.
The pitcher-specific continuous decline parameters and batter-specific discontinuity parameters have Gaussian priors,
\begin{align}
\label{eqn:priors_playerSpecificSlopes}
\begin{cases}
    \alpha_{0kp(i)} \sim \N(\alpha_{0k}, \sigma^2_{0k}), \\
    \alpha_{1kp(i)} \sim \N(\alpha_{1k}, \sigma^2_{1k}), \\
    \beta_{2kb(i)} \sim \N(\beta_{2k}, \sigma^2_{2k}), \\
    \beta_{3kb(i)} \sim \N(\beta_{3k}, \sigma^2_{3k}),
\end{cases}
\end{align}
which themselves have priors,
\begin{align}
\label{eqn:priors_playerSpecificSlopes_2}
\begin{cases}
    \alpha_{0k}, \alpha_{1k}, \beta_{2k}, \beta_{3k} \sim \N(0, 25), \\
    \sigma^2_{0k}, \sigma^2_{1k}, \sigma^2_{2k}, \sigma^2_{3k} \sim \text{half } \N(0,1).
\end{cases}
\end{align}
% Note that in this model, we don't force a positive slope the pitcher-specific parameter reflecting continuous change across a game to have .

\begin{figure}[H]
\centering
\includegraphics[width=0.7\textwidth]{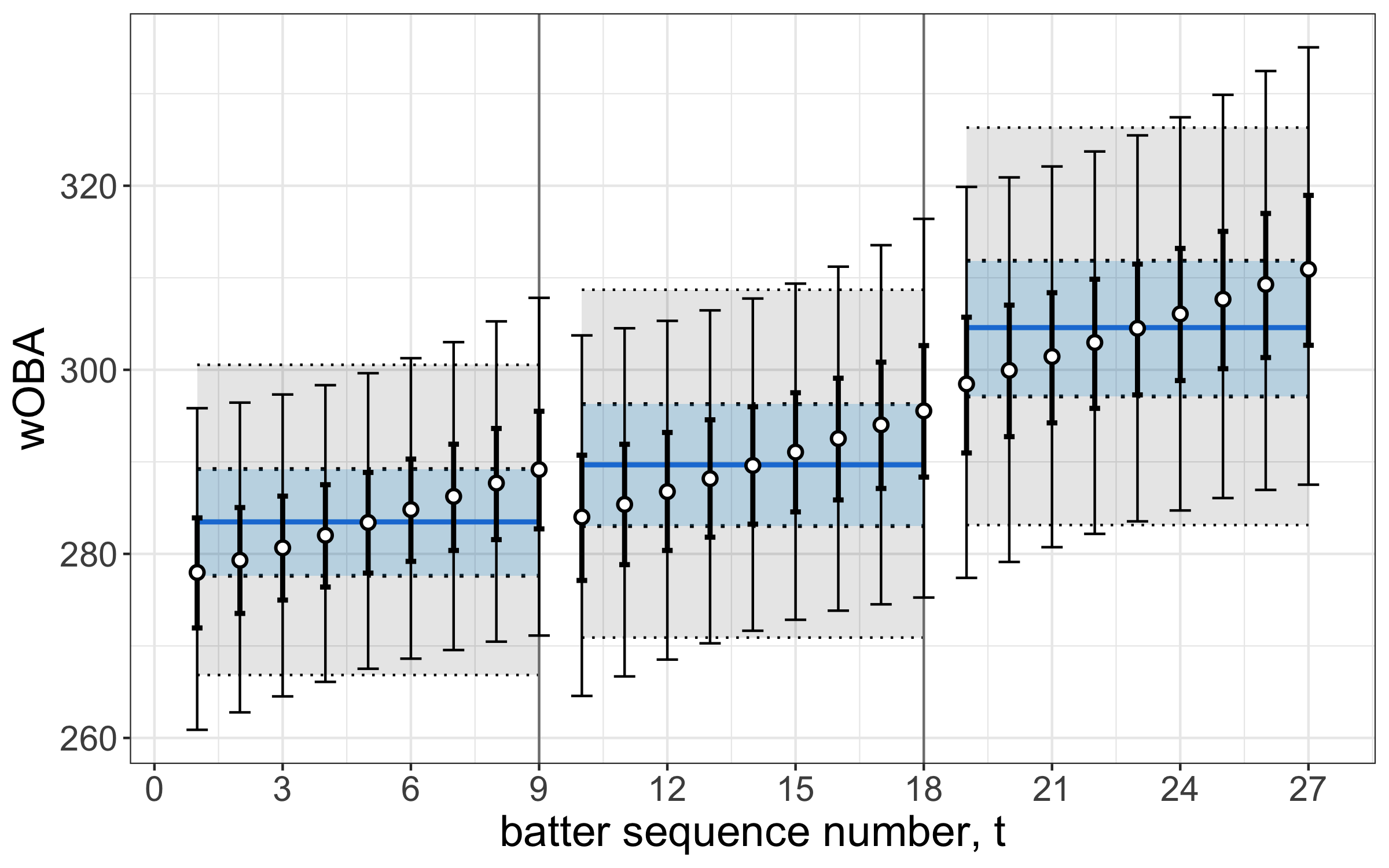}
\caption{
Trend in expected $\woba$ over the course of a game in 2017 for an average batter facing an average pitcher of the same handedness on the road, according to the model from Equation~\eqref{eqn:model_playerSpecificSlopes}.
The white dots indicate the posterior means of the expected $\woba$ values, the thick black error bars denote the 50\% credible intervals, and the thin black error bars denote the 95\% credible intervals.} 
\label{fig:xwoba_trend_1_2017_pitVaryingModel}
\end{figure}

With this more flexible model, the qualitative results of our study don't change.
For instance, as in Figure~\ref{fig:plot_xwoba_over_time_2017a}, in Figure~\ref{fig:xwoba_trend_1_2017_pitVaryingModel} we plot the posterior distribution of the trajectory of expected $\woba$ over the course of a game, according to the player-specific model from Equation~\eqref{eqn:model_playerSpecificSlopes} fit on data from 2017.
In particular, we use the posterior distributions of the prior means $\alpha_{0k}$, $\alpha_{1k}$, $\beta_{2k}$, and $\beta_{3k}$ to compute the $\xwoba$ trajectory for an average pitcher facing an average batter.
% We do not see the exact same $\xwoba$ trajectory as before because our model is different now.
We do not see a significant upwards discontinuity in expected $\woba$ from one $\tto$ to the next.
In other words, we find little evidence for a strong batter discontinuity between times through the order.
This trend is persistent across each year from 2012 to 2019.

%%%%%%%%%%%%%%%
%%%%%%%%%%%%%%%

\end{document}